\documentclass[preprint]{aastex}
\def \sun {$_{\scriptscriptstyle \odot}$}

\shorttitle{3-Dimensional Core-Collapse}
\shortauthors{Fryer \& Warren}

\received{2003 June 6}
\begin{document}

\title{The Collapse of Rotating Massive Stars in 3-dimensions}


\author{Chris L. Fryer and Michael S. Warren}
\affil{Theoretical Astrophysics, Los Alamos National Laboratory, 
Los Alamos, NM 87545}

\begin{abstract}

Most simulations of the core-collapse of massive stars have focused on
the collapse of spherically symmetric objects.  If these stars are
rotating, this symmetry is broken, opening up a number of effects that
are just now being studied.  The list of proposed effects span a range
of extremes: from fragmentation of the collapsed iron core to
modifications of the convective instabilities above the core; from the
generation of strong magnetic fields which then drive the supernova
explosion to the late-time formation of magnetic fields to produce
magnetars after the launch of the supernova explosion.  The list of
observational effects of rotation ranges from modifications in the
gamma-ray line spectra, nucleosynthetic yields and shape of supernova
remnants caused by rotation-induced asymmetric explosions to strong 
pulsar radiation, the emission of gravitational waves, and altered 
r-process nucleosynthetic yields caused by fast-spinning rotating 
stars.

In this paper, we present the results of 3-dimensional collapse
simulations of rotating stars for a range of stellar progenitors.  We
find that for the fastest spinning stars, rotation does indeed modify
the convection above the proto-neutron star, but it is not fast enough
to cause core fragmentation.  Similarly, although strong magnetic
fields can be produced once the proto-neutron star cools and
contracts, the proto-neutron star is not spinning fast enough to
generate strong magnetic fields quickly after collapse and, for 
our simulations, magnetic fields will not dominate the supernova explosion
mechanism.  Even so, the resulting pulsars for our fastest rotating
models may emit enough energy to dominate the total explosion energy
of the supernova.  However, more recent stellar models predict
rotation rates that are much too slow to affect the explosion, but
these models are not sophisticated enough to 
determine whether the most recent, or past, stellar rotation rates are
most likely.  Thus, we must rely upon observational constraints to
determine the true rotation rates of stellar cores just before
collapse.  We conclude with a discussion of the possible constraints on
stellar rotation which we can derive from core-collapse supernovae.

\end{abstract}


\section{Introduction}

The collapse of a massive star to a proto-neutron star releases an
enormous amount ($\sim 10^{53}$\, ergs) of gravitational energy.  This
energy is primarily converted into thermal energy and later radiated
away in the form of neutrinos.  It is believed that some of this
neutrino energy will be deposited in the star and drive a strong
supernova explosion and multi-dimensional models with simplified
neutrino transport have produced explosion (e.g. Burrows, Hayes, \&
Fryxell 1995; Fryer 1999; and references therein).  These successful
explosions eject far too much neutron-rich material to match the
observed nucleosynthesis production.  Models with the most
sophisticated transport to date (Bruenn, DeNisco, \& Mezzacappa 2001;
Liebend\"orfer et.  al. 2001; Buras et al. 2003; and references
therein) hover on the line between supernova explosion and direct
collapse into a black hole and the collapse models with the best
neutrino neutrino transport to date do not produce explosions with
current stellar progenitors.  It appears that the neutrino driven
supernova mechanism depends on the detailed implementation of the 
physics involved (from neutrino transport to the equation of 
state for matter and the effects of general relativity).

Although there is considerable evidence that the supernova mechanism
is not robust\footnote{Since roughly 10-40\% of stellar collapses
  result in the formation of black holes (Fryer \& Kalogera 2001), it
  appears that nature also finds that the supernova mechanism depends
  upon the details of the collapsing star (mass, rotation, ...).}, the
the fact that the neutrino-driven supernova mechanism is not robust
has led some astronomers to search for alternate mechanisms to produce
supernova explosions.  If the star is rotating, a sizable amount of
the energy released can be converted into rotational energy.  LeBlanc
\& Wilson (1970) proposed that magnetic fields could extract this
rotational energy and produce a strong supernova explosion, but when
M\"uller \& Hillebrandt (1981) found that large magnetic fields were
required for such a mechanism to work, interest in this mechanism
declined.  Even if magnetic fields are not the driving force behind
core-collapse supernovae, rotation and magnetic field effects may play
an important role in supernova explosions.

Rotation has still been studied in core collapse, both as a
means to produce gravitational waves (see Dimmelmeier, Font, \&
M\"uller 2002; Fryer, Holz, \& Hughes 2002; New 2003 and
references therein) and as a means to produce asymmetric explosions
(see Fryer \& Heger 2000 - hereafter FH - and references therein).
FH found that, at least in 2-dimensional models, rotation
weakened the convection in the rotating plane, ultimately leading to
explosions that are strongest along the rotation axis.  However, the
symmetry axis of these 2-dimensional simulations lay along the
rotation axis, and it is difficult to distinguish numerical boundary
effects from the true asymmetry in the explosion.  The first goal of
this paper is to test the conclusions of FH by
running a series of 3-dimensional, fully $4\pi$ (no boundaries),
models of the collapse of rotating stars.

However, rotation may play a larger role in producing explosions for
some systems.  With the discovery of gamma-ray bursts and hypernovae,
rotation has once again been invoked as a source of explosion energy
(Woosley 1993; H\"oflich et al 1996; Iwamoto et al. 1998).  In the
case of the collapsar GRB model and hypernovae, it is believed that
the explosion is driven after the the core has collapsed to a black
hole and the explosion engine can take advantage of the enormous
rotational energy held by the rapidly rotating outer layers of the
star.  But some authors (e.g. Akiyama et al. 2003) believe magnetic 
field generation is efficient enough for rotation to drive an 
explosion quickly and leave behind a neutron star.
We will test this possibility with currently available stellar models.

The level of asymmetry in the explosions not only tests the nature of
the collapse model, but also the progenitor itself.  Unfortunately,
the rotation rate of a star prior to collapse depends sensitively on
the recipes used to model angular momentum transport in 1-dimensional
stellar evolution codes.  On the other hand, interpretations of
supernova observations have not yet arrived at a consensus on the
level of asymmetry in the explosion: some groups insist large
asymmetries are required (Wang et al. 2002 and references therein)
while others argue mild asymmetries are sufficient (Nagataki 2000;
Hungerford, Fryer \& Warren 2003).  Better models of the explosion and
increased supernova observations will be able to distinguish the level
of asymmetry, and hence magnitude of rotation, in the collapse and
explosion of massive stars.  However, other tests can be, and have
been, used to measure the rotation in the core: pulsar spin rates,
gravitational radiation, and nuclear yields.  We conclude with a
discussion of these potential observational tests of core-collapse
rotation.

The outline of this paper is as follows: \S 2 describes our numerical 
technique and the progenitors used in our simulations, \S 3 concentrates 
on the effects of rotation on the standard neutrino-driven supernova 
mechanism, \S 4 details the viability of additional effects of 
rotation on the supernova explosion mechanism beyond the basic 
neutrino-driven picture: from fragmentation to the magnetic-field 
supernova mechanism.  We conclude with a discussion of the observational 
signatures from these rotationally-induced modifications to the supernova 
explosion with an eye towards constraining the rotation of the iron 
core of a pre-collapse star.

\section{Progenitors and Numerical Techniques}

For these simulations, we use the 3-dimensional, smooth particle
hydrodynamics (SPH) supernova code discussed in Fryer \& Warren
(2002), Warren et al. (2003).  Because these simulations involve
rotating progenitors where gravity is less likely to be symmetric, we
do not constrain the calculation to spherical gravity, but instead use
the tree-based gravity algorithm described in Warren \& Salmon
(1993,1995), Warren et al. (2003) to calculate the multi-dimensional
effects of gravity.  

The neutrino transport and equation of state physics uses the same
algorithms described in Herant et al. (1994) and Fryer (1999) for
their 2-dimensional results.  There are 2 major differences between
these implementations and those of current 2-dimensional simulations.
First, we use a single-energy, flux-limited diffusion algorithm to
transport the neutrinos (although we do include transport for each of
3 neutrino species - electron, anti-electron, and $\mu + \tau$
neutrinos).  Mezzacappa \& Bruenn (1993) found that such a simplified
transport scheme overestimated the total energy deposition by
neutrinos.  This is the primary explanation why the more sophisticated
treatments in 2-dimensions (e.g. Buras et al. 2003) have not produced
explosions whereas those single-energy flux-limited diffusion (e.g.
Fryer 1999) succeed.  The single-energy flux-limited diffusion scheme 
used in this paper will tend to produce explosions more easily than 
those simulations using more sophisticated neutrino transport.

However, equation of state problems also alter the results.  Most
codes use the Swesty-Lattimer equation of state (Lattimer \& Swesty
1991) to explain the behavior of matter not only at high densities but
also down through nuclear statistical equilibrium.  An error in this
equation of state (Lattimer - pvt. communication) leads to incorrect
estimates of the entropy just after bounce and slower growth in the
initial convection.  Our technique circumvented this issue by limiting
the use of the Swesty-Lattimer equation of state to densities above
$10^{11} {\rm g cm^{-3}}$ where the effects of this error were
minimal and implementing a nuclear statistical equilibrium network 
in the lower-density regime.  Our corrections to the equation of state 
led to more vigorous convection early on and helped to produce explosions.  

These 2 effects make it more likely for our simulations to produce
explosions.  Although the fix to the equation of state is definitely
an improvement upon what groups like Buras et al. (2003) use, we must
temper our results by the fact that our simplified neutrino transport
may produce artificial explosions.  Hence, we will not focus on the
explosions produced in these simulations, but on the effects of
rotation on the convection during collapse and on the nature of the
final remnant after collapse.  These simulations provide a
3-dimensional probe of the effects of convection from which we can
start to build up our intuition on this wrinkle in the core-collapse
problem.  Rotation adds additional numerical uncertainties into our
simulations.  Shear forces become more important with our rotating
models, and we will concentrate our discussion on the numerical
artifacts of this shear.

First, let us discuss our progenitors.  To compare with the results of
FH, we use the same standard progenitor used by this work: E15B of
Heger, Langer \& Woosley (2000).  To study the effects of different
progenitors, we also use model E15A of Heger, Langer \& Woosley (2000)
and the slowly rotating model 15\,M\sun of Heger, Woosley, \& Spruit
(2003), corresponding to models SN15A, SN15B, and SN15C respectively
in this work (Table 1).  Although these models have a range of angular
velocities (Fig. 1)\footnote{FH used only model E15B in their
simulations.  Note, however, that they erroneously plotted the angular
momentum of SN15A in Fig. 3 of their paper.}, they have much lower
total angular momenta than what is assumed in most gravitational wave
simulations.  As a calibration for rotation, we also have run models
E15A and E15B with the angular velocity set to zero (models SN15A-nr,
SN15B-nr).

We use the angular and radial velocity distribution given in Heger et
al. (2000) to determine the 3-dimensional velocity vectors onto our
initial distribution of particles.  Like the Fryer \& Warren work, we
construct our 3-dimensional star (filling the full $4 \pi$ star) by
setting up a series of radially spaced shells (each filled with
randomly, but roughly equally, spaced particles).  The separation
between shells is set to the mean separation between particles within
each shell.  This random setup, although it prevents any artifacts
based on grid issues, does lead to some density deviations in our
initial model.  These deviations are largest at composition boundaries
(e.g. silicon and oxygen shell boundaries), and do not exceed 10\% of
the density (in most of the star, the deviation is less than 5\%) for a 1
million particle initial model.  These variations are on par with the
deviations expected from silicon and oxygen flash burning (Bazan \&
Arnett 1998), so although they are numerical in origin, they may match
nature reasonably well.

An important constraint on our models is the numerical shear produced
by the artificial viscosity used in most SPH codes (e.g. Benz 1990).
This viscosity is generally implemented with the following form:
\begin{equation}
\Pi_{ab} \ = \ \left \lbrace 
\begin{array}{ll}
\displaystyle{\frac{-\alpha c_{ij} \mu_{ij} + \beta \mu_{ij}^2}
{\rho_{ij}}} & {\rm if} \mu_{ij} < 0; \\
                          0        & {\rm otherwise.} 
                     \end{array}
\right.
\end{equation}
Here, $\mu_{ij} = h (\vec{v}_{i}-\vec{v}_{j}) \cdot
(\vec{r}_i-\vec{r}_j)/(|\vec{r}_i-\vec{r}_j|^2 + \epsilon h^2)$ is the
velocity divergence at particle $i$ due to particle $j$ where
$\vec{v}_{i},\vec{v}_{j}$ are velocities of particles $i,j$
respectively, $\vec{r}_{i},\vec{r}_{j}$ are velocities of particles
$i,j$ respectively, $\epsilon$ is a small offset to avoid numerical
artifacts from particles very near each other, and $h$, $c_{ij}$ and
$\rho_{ij}$ are the average smoothing length, sound speed and density
of particles $i$ and $j$.  $\alpha$ and $\beta$ are artificial
viscosity parameters.  Most of our models use this implementation of the
artificial viscosity with the standard values for the viscosity terms:
$\beta = 2 \times \alpha = 3.0$.  However, to test the effects of this
viscosity on the angular momentum transport, we have also run one
model (SN15A-rv) with these viscous terms reduced by a factor of 10
($\beta = 2 \times \alpha = 0.3$).

Our collapse simulations model the inner 3\,M\sun of the collapsing star
with 1 million SPH particles.  The resolution is increased near the
entropy-driven convective region, corresponding to an angular resolution 
of over 20 particles per square degree and a mass resolution better than 
$10^{-6}$\,M\sun per particle.  We have run one 5 million particle model 
using the E15A progenitor (SN15A-hr).  The qualitative convection features 
in this high resolution run did not differ considerably from the 
lower resolution simulations.

\section{Rotational Effects on Collapse and Explosion}

For the rotation rates found in modern supernova progenitors, rotation
has a limited set of effects on the collapse of a massive star.  The
primary effects of rotation (see Shimizu, Yamada, \& Sato 1994; Kotake
et al. 2003; FH and references therein) are: 1) high angular velocity
material does not accelerate as much in collapse, leading to weaker
bounces and lower entropies for this material, 2) angular momentum
limits convection where angular momentum is highest and 3) the
deformation in the neutrinosphere causes an asymmetry in the neutrino
emission.  In this section, we will first concentrate on the purely
hydrodynamic effects which FH argued to be the dominate rotation
effects.  We will then discuss the effects of rotation on the neutrino
emission.  As we shall show, both effects occur in rotating modles and 
can contribute to asymmetries in the explosion.  The pure hydrodynamic 
effects occur immediately and are most important for quick explosions 
($<100$\,ms after bounce) which is the case for our simulations, but neutrino 
asymmetries will play a larger role if the explosion is more delayed.

\subsection{Hydrodynamic Effects of Rotation}

In this paper, we will use the term ``polar'' material for that matter
which lies along the axis or pole of the rotation and ``equatorial''
material for that material which lies in the plane of the rotation.
The equatorial material has the bulk of the angular momentum in the
collapsing star and it is this material that slows during collapse,
resulting in a weaker bounce and lower shock values.  It is also in
the equator where the angular momentum profile strongly inhibits
convection.

The collapse of a massive star occurs when the temperature and density
in the iron core become high enough to dissociate the iron (removing
thermal pressure) and to induce electron capture (removing electron
degeneracy pressure) in the core.  The initial compression induces
further electron capture and iron dissociation resulting in a runaway
collapse that halts only when the core reaches nuclear densities where
nuclear forces cause the core to ``bounce''.  This bounce sends a
shock through the star which heats the star, raising its entropy.
When the shock stalls (through neutrino and dissociation energy
losses), it leaves behind an unstable entropy profile that seeds the
convection.  This convection is currently believed to be important in
the success of the supernova mechanism (Herant et al. 1994; Burrows,
Hayes, \& Fryxell 1995; Janka \& M\"uller 1996; Fryer 1999; Rampp \&
Janka 2002).  With their 2-dimensional simulations, FH argued that the
combined effect of the lower entropy from weaker shocks and the strong
angular momentum gradients in the equatorial region of the star led to
less convection in this region, and ultimately, weaker explosions
along the equator versus the rotation axis.  Indeed, FH found that the
explosion velocities along the pole were twice as fast as those along
the equator.  Such asymmetric explosions aid in the outward mixing of
the nuclear burning products of the supernova (Nagataki 2000;
Hungerford, Fryer, \& Warren 2003).  In this section, we review the
claims of FH to determine whether such strong asymmetries persist in
3-dimensional simulations.

First, let's address the first point from FH: does the polar region
reach higher entropies after bounce due to a stronger bounce in the
pole?  Fig. 2 shows an angular slice of model SN15A-hr just before
bounce.  The colors show the radial velocity and the vectors denote
the direction and magnitude of the velocity.  Along the poles, where
the centrifugal support is negligible, the velocities are much higher
than in the equator.  The velocity of the infalling material is 
approximated by assuming free-fall conditions:
\begin{equation}
v_{\rm infall} \approx \sqrt{\frac{G M_{\rm enclosed}}{r} - \frac{j^2}{r^2}},
\end{equation}
where $G$ is the gravitational constant, $M_{\rm enclosed}$ is the
enclosed mass, $j$ is the specific angular momentum of the infalling
material, and $r$ is its collapse radius.  For the angular momentum in
SN15A, the corresponding polar vs. equatorial infall velocities as the
1.2M\sun shell reaches 75\,km are: 6.5,3.8$\times 10^9 {\rm cm \,
s^{-1}}$ pole vs. equator respectively.  When the shell reaches 50\,km,
the respective infall velocities are: 8$\times 10^9,0 \rm{(centrifugal 
hang-up)} {\rm cm \,
s^{-1}}$ pole vs. equator.  The equatorial material simply does not
reach the high infall velocity magnitudes that are achieved along the
pole.  The infalling material stops abruptly when the core
``bounces'', causing it to shock.  The entropy (S) in units of
Boltzman's constant per nucleon for such a strong shock (assuming a
polytropic equation of state with $\gamma=4/3$) is:
\begin{equation}
S \approx 9.9 \rho_9^{-1/4} v_{10}^{3/2} {\rm k_{B} \, per \, nucleon},
\end{equation}
where $\rho_9$ is the density and $v_{10}$ the velocity of the
infalling material in units of $10^9{\rm \,g\,cm^{-3}}, 10^{10}{\rm
  \,cm\,s^{-1}}$ respectively.  The faster moving material along the
poles has a stronger shock and its resultant entropy after bounce is
greater.  Hence, after bounce, we would expect to see higher entropy
material along the rotation axis.  Fig. 3 shows the constant entropy
isosurfaces for model SN15A-hr 45\,ms after bounce.  The isosurface
for an entropy of 7.5${\rm k_{B}}$ per nucleon is slightly elongated
along the pole, but exists both in the equatorial and polar regions.
By studying increasingly higher entropy isosurfaces (Fig. 3 shows
entropy isosurfaces of 7.5, 8.0, and 9.0 ${\rm k_{B}}$ per nucleon),
we see that the highest entropies are limited to the polar region.  We
remind the reader that our 3-dimensional simulations model the entire
$4 \pi$ sphere with a randomly distributed shell setup, so these
asymmetries can {\it not} be boundary effects or artifacts of the
initial conditions.

The stall of the bounce shock leaves behind a negative entropy
gradient which initiates convection above the surface of the
proto-neutron star.  Due to the higher entropy where the shock
velocities were largest, this gradient is much larger along the poles,
driving stronger convection in this region.  When rotation is
included, a negative entropy gradient is not sufficient to drive
convection.  As FH pointed out, supernova convection is inhibited by
the large angular momentum gradient in the poles.  For rotating
objects where the angular momentum increases with radius, convection
occurs only if the negative entropy gradient can overcome the positive
angular momentum gradient (also known as the Solberg-H{\o}iland
instability criterion - Endal \& Sofia 1978):
\begin{equation}
\frac{g}{\rho} \left [ \left ( \frac{d \rho}{d r} \right
)_{\rm adiabat} - \frac{d \rho}{d r} \right ] -
\frac{1}{r^3} \frac{d j^2}{d r} > 0,
\end{equation}
where $g=G M_{\rm enclosed}/r^2$ is the gravitational acceleration
with gravitational constant $G$ and $M_{\rm enclosed}$, $\rho$, and
$j$, the enclosed mass, density and specific angular momentum
respectively at radius r in the collapsed core.  Fig. 4 shows the
angular momentum part of equation 4 ($1/r^3 \, d j^2/dr$) for models
SN15B (our fastest rotating progenitor) vs. SN15C (our slowest
rotating progenitor) both along the equator and pole.  Note that along
the equator, this value is 4 orders of magnitude higher for the fast
rotating vs. slowly rotating models.  Indeed, the constraint is
essentially negligible in the poles of both models and even in the
equator of the slowly-rotating model.  Only models SN15A and SN15B are
rotating fast enough that convection is affected by the rotation rate.
Fig. 5 shows the upflows (isosurfaces of material moving with outward
radial velocities of 3000\,km\,$s^{-1}$) for many of our models 25\,ms
after bounce.  For our rotating models (SN15A,SN15A-hr,SN15B), the
convection is strongest along the poles, but for our non-rotating and
slowly rotating models (SN15A-nr,SN15B-nr,SN15C), the convection has
no preferred axis.  We have thus confirmed the FH claim that
convection is strongest in the poles for sufficiently rapidly rotating
progenitors.  Note, however, the slowly rotating progenitor from Heger
et al. (2003) is not rotating fast enough to effect the convection,
and hence, the explosion through the neutrino-driven supernova
mechanism.

\subsection{Neutrino and Explosion Asymmetries from Rotation}

Shimizu et al. (1994) and Kotake et al. (2003) have argued that due to
the asymmetry in the neutrinosphere, asymmetric neutrino heating will
also drive stronger convection along the poles.  Density isosurfaces
($10^{11},10^{12} {\rm g \, cm^{-3}}$) for model SN15A-hr are shown in
Fig. 6.  Even the inner $10^{12} {\rm g \, cm^{-3}}$ density
isosurface has axis ratios of roughly 1.4:1 equator vs. pole.  This
asymmetry leads to an asymmetry in the neutrinosphere, and ultimately
in the neutrino heating which Shimizu et al. (1994) and Kotake et
al. (2003) have argued will drive a jet-like explosion if the
explosion mechanism is sufficiently delayed.  However, for model
SN15A-hr 45\,ms after bounce, the asymmetry in the neutrino heating is
so small that the timescale for the asymmetry in neutrino heating to
alter the convection ($t_{\rm heating} = (u_{\rm pole}-u_{\rm
equator})/ (du_{\rm pole}-du_{\rm equator})$ where $u_{\rm pole,
equator}$ is the internal energy and $du_{\rm pole, equator}$ is the
neutrino heating for the poles and equator) is over 100\,ms.  The 
delay must be at least long to see the effects of rotational heating.

In our simulations, however, the shock moves out before such
asymmetric heating can make a large difference in the convection, and
it is more likely that the asymmetry is entirely due to the purely
hydrodynamic effects listed by FH.  To determine whether such
asymmetric convection yields asymmetric explosions, we must model the
propogation of the expanding shock until it develops into an
explosion.  In 2-dimensions, FH were able to model the explosion
1.5\,s past bounce, giving time for the expanding material to develop
into an explosion with over 20\% of the total explosion energy already
converted into kinetic energy.  A 
1.5\,s post-bounce simulation is beyond our computational resources 
for our 3-dimensiona, flux-limited diffusion code.
Fig. 7 shows the upflows (isosurfaces of material moving with outward
radial velocities of 2000\,km\,s$^{-1}$) of model SN15A at a series of
times past bounce.  From Fig. 7, it is clear that as the explosion
develops, it remains strongest along the poles and the expanding shock
moves out furthest in this direction.  It is likely that the resultant
explosion will be strongly asymmetric along the poles.  Although we
can not quantitatively estimate the level of asymmetry from our
simulations, given the excellent agreement in the convection with FH,
it is not unreasonable to assume that the mean velocity of the ejecta
along the poles could be a factor of 2 (or greater) larger than the
velocity along the equator in these rapidly rotating explosions (see
FH).

Recall that our simplified neutrino transport (\S 2) may cause the 
explosion to occur faster than reality and the asymmetries 
in the neutrino heating can be important.  If this case is true, 
we would also expect an explosion which is strongest along the 
rotation axis (Shimizu et al. 1994 and Kotake et al. 2003).

\section{Other Rotational Effects on the Explosion}

As these rotating models collapse, much of the potential energy 
can be converted into rotational kinetic energy and the rotational 
energy can exceed the supernova explosion energy.  This energy may 
be tapped to help drive the supernova explosion.  If the rotational 
energy gets high enough, it can cause the core to go unstable and 
fragment.  Strong magnetic fields may also develop quickly and extract 
the rotational energy to drive the supernova explosion.  At late 
phases, as the core contracts, this rotational energy increases, 
producing rapidly spinning neutron stars that, if born with strong 
enough magnetic fields, can be observed as pulsars.

What are the implications for the rotating progenitor stars modeled in
this paper?  To know this, we must first understand the the evolution
of the angular momentum distribution in the collapsing core.  As one
might expect, most of the angular momentum is concentrated along the
equator after collapse in an ellipsoidal core (Fig. 8).  Fig. 9 shows
the angular momentum in mass zones for 3 separate time slices during
the collapse.  As the proto-neutron star contracts, the angular
momentum is gradually transported out of the central core.  For SPH,
angular momentum is explicitly conserved (Benz 1990) and angular
momentum losses are limited to round-off errors (roughly one part in a
million for our simulations).  However, angular momentum can be
transported artificially in these simulations.  To test whether the
transport of angular momentum in Fig. 9 was numerical or real, we ran
2 additional models: a high resolution run (SN15A-hr) where the
effects of artificial SPH viscosity should be diminished, and a run
where the SPH viscosity is reduced by a factor of 10 (SN15A-rv).  If
the effects of numerical viscosity were indeed the culprit behind the
angular momentum transport, the high-resolution and reduced viscosity
simulations should have yielded quite different results from model
SN15A.  However, as we see from the angular velocity distribution in
Fig. 10, there is very little difference between these 3 rotating
simulations.

How can we explain the outward movement (in mass zones) of the angular
momentum?  As the core contracts, the material with greater angular
momentum is slowed by centrifugal forces.  It does not collapse as
deeply as the low angular momentum material and piles up at higher
radii where we find much of the angular momentum deposited.  Fig. 11
shows the distribution of material in model SN15A 90\,ms before bounce
which ultimately contracts to form the proto-neutron star.  A large
fraction of the material below the isocontour boundaries collapses down
to 12.5, 20, 50\,km 140\,ms after bounce.  The actual fraction is
denoted by solid (65\%) and dotted (95\%) lines.  Figure 11 tells us
what part of the original star collapses within 12.5, 20, 50\,km and
ultimately forms the neutron star.  Material in the equator does not
contract so quickly and does not initially become part of the neutron
star and the neutron star mass is originally biased toward the low
angular momentum material of the polar region.  This is not to say
that the high angular momentum material will not ultimately become
part of the neutron star.  Much of this material will
slowly find its way on to the proto-neutron star as it either cools
and loses pressure support or sheds some of its angular momentum along
with some mass, losing the support of centrifugal forces.  

Note that if we had assumed that the angular momentum was conserved 
with mass down through collapse, we would have overestimated the 
angular momentum in the inner core by roughly a factor of 5 (and 
hence the total rotational energy by over an order of magnitude).  
With this revised understanding of the angular momentum distribution, 
let's study the additional rotational effects on the supernova explosion.

\subsection{Core Fragmentation}

There has been a growing number of recent papers reviving the idea
that fragmentation can occur in the collapse of massive stars (Fryer,
Woosley, \& Heger 2000; Davies et al. 2002; Colpi \& Wasserman 2002).
However, the conditions to cause fragmentation are extreme, requiring
rotation rates that can hang up the infalling material, causing the
density to peak, not in the center, but in a torus.  Even with these
conditions, many simulations find that the torus then forms a bar (not
separate fragments) and ejects its angular momentum to coalesce into a
single compact object (e.g. Centrella et al. 2001).

Even our fastest rotating cores do not lead to fragmentation of the
core.  The nature of rotating, gravitationally-bound objects is
fairly well determined by the ratio of rotational energy over the
magnitude of the gravitational binding energy:  $T/|W| = 0.5 I_{\rm
core} \Omega_{\rm core}^2/ (G M^2_{\rm core}/R_{\rm core})$.  Here
$I_{\rm core}, \Omega_{\rm core}, M_{\rm core}, R_{\rm core}$ are,
respectively, the collapsed core's moment of inertia, angular
velocity, mass, and radius and $G$ is the gravitational constant).
This ratio as a function of enclosed core mass for our 3 rotating
progenitors (SN15A, SN15B, and SN15C) 45\, ms after bounce is plotted
in Fig. 12.  If the density is centrally peaked, this ratio must
exceed $\sim 0.14$ for the core to be secularly unstable.  Given that
the angular momentum can be transported outward, dynamical
instabilities ($T/|W|>0.25$) are probably required to produce bar
instabilities and fragmentation.  Note that only our fastest rotating
star (SN15B) has values even close to this secular instability.  It is
therefore unlikely that any of these models will develop bar modes,
let alone fragment, at these early times.  Not surprisingly, bar 
modes do {\it not} develop in any of our simulations and, likewise, 
fragmentation is not an issue.

Indeed, even if the star is dynamically unstable, fragmentation is
most likely when the density profiles are not centrally peaked.  Such
toroidal density profiles are more likely to become dynamically
unstable (Centrella et al. 2002 and references therein) and could then
more easily fragment.  Note, however, that the Centrella et al. (2002)
simulations did not show fragmentation even though the star was
dynamically unstable and toroidal density distributions are probably a
necessary, but not sufficient, condition for fragmentation.  With our
more moderate rotation speeds, we would expect ellipsoidal density
configurations such as Maclaurin spheroids.  Our simulations did relax
into these aspherical density profiles (Fig. 6).  With these centrally
peaked density profiles, it is extremely unlikely that fragmentation
or dynamical bar instabilities will occur with any of the
currently-produced supernova progenitors.  Despite the recent burst of
papers on fragmentation in core-collapse, it is almost certain that
fragmentation will only occur in rare collapse cases where the star
has been spun up prior to collapse (e.g. in Fryer et al.  2001, the
collapsed ``core'' consisted of the central $\sim 50$\,M\sun of a
collapsing 300M\sun star, the outer layers of which contained a great
deal of angular momentum).  If the current collapse progenitors are
typical for supernova progenitors, most collapsing stars will not
fragment.

\subsection{Magnetic Field Driven Explosions}

Even with the redistribution of angular momentum, as the core
contracts, the rotational energy in the star increases.  A number of
mechanisms exist in which rapidly rotating proto-neutron stars can
generate magnetic fields (e.g. Thompson \& Duncan 1993, Akiyama et al.
2003).  Whether or not these magnetic fields can play the dominant
role in driving the explosion relies upon how quickly such magnetic
fields can develop (if an explosion is not launched quickly - within
the $\sim$1st second after bounce - the proto-neutron neutron star will
accrete too much material and collapse into a black hole).  The dynamo
proposed by Thompson \& Duncan (1993) is unlikely to make a strong magnetic
field until the neutron star cools, contracts, and spins up.  This 
dynamo can easily make magnetar fields in the cooling neutron star 
after the launch of the explosion, but Fryer \& Heger (2000)
found that this dynamo could not make strong enough fields to effect 
the convective region in the first second after bounce.  It is in 
this first second that the star either explodes or collapses down to 
a magnetic field and it is thus unlikely that this dynamo mechanism 
will produce the initial supernova explosion.

Akiyama et al. (2003) argue that a very effective magnetic field process can
arise from the differential rotation in the core.  Once these magnetic
fields form, they argue that these magnetic fields can extract
rotational kinetic energy to drive a supernova explosion.  The 
saturation magnetic field $B$ arising from differential rotation 
is given by (Akiyama et al. 2003):
\begin{equation}
B^2 \sim 4 \pi \rho r^2 \Omega^2 (d {\rm ln} \Omega/ d {\rm ln} r)^2
\end{equation}
where $\rho$ is the density and $\Omega$ the angular velocity at a
radius $r$ in the proto-neutron star.  Even for our most rapidly
rotating progenitors, it is difficult to produce magnetic fields in
excess of $10^{14}$G (Fig.13).  Although the rotational energy in our
fastest rotating cores can exceed $10^{51}$\,erg (Tab. 1), the
estimated magnetic fields are too weak to effectively use this energy
and dominate the explosion.  Even so, we can not rule out that
magnetic fields can't play some role driving further asymmetries in
the ejecta.  For our slow rotating (SN15A) progenitor, magnetic field
driven effects will not become important until after the proto-neutron
star cools and contracts, the total rotational kinetic energy in the
proto-neutron star (and in the entire core for that matter) is
insufficient to produce a $10^{51}$\,erg explosion.

For our models, the currently proposed mechanisms to increase the
magnetic field strength do not yield field strengths that are strong
enough to qualitatively alter the explosion dynamics.  In a delayed
explosion explosion, the core would compress somewhat, allowing for
higher rotation speeds and possibly higher differential rotation.  In
our highest rotating progenitors, this compression may be sufficient
to allow magnetic fields to play a more dominant role in the
explosion.  Hence, we can not rule out that magnetic fields won't play
any role on the explosion.  Of course, if the explosion were to be
launched several seconds after the bounce of the core (when the star
has collapsed to a black hole), there would be ample magnetic fields
and rotational energy to drive an explosion for our rapidly rotating
progenitors (SN15A, SN15B).  Such an explosion mechanism is known as a
collapsar (Woosley 1993) and has been suggested as a gamma-ray burst
and hypernova engine.

\subsection{Pulsar Emission and the Supernova Explosion}

Another way that magnetic fields can effect the explosion is through
intense pulsar emission shortly after the explosion.  A millisecond
pulsar with a strong magnetic field could easily inject a $10^{51}$
erg jet into the expanding supernova ejecta.  Such a jet would
significantly alter the observed supernova spectra, polarization and
light curve.  Here we discuss the expected spin periods from our
simulations and the magnetic fields required to cause the pulsar
emission to dominate the supernova light curve.

If we take the total angular momentum in the 1.0\,M\sun proto-neutron
star at the end of our simulations and conserve it as it contracts
into a 12.5\,km neutron star, model SN15B would produce a a
sub-millisecond pulsar.  However, beware of such calculations.  We are
estimating the spin period by assuming that the angular momentum will
be conserved as the proto-neutron star collapses.  For models SN15A
and SN15C, where we have late-time calculations, we find that the
inner core continues to lose angular momentum.  45\,ms after bounce,
if we took the inner core of model SN15C and collapsed it, conserving
angular momentum, we would find that the neutron star would have a
spin of 12\,ms.  90\,ms later, the collapsing core had only enough
angular momentum to produce a 17\,ms pulsar.  It is likely that the
spin period of this slow rotating model will ultimately be greater
than 20\,ms.  

Why does this occur?  As we showed in Fig. 8, much of the high angular
 momentum material does not collapse directly into the proto-neutron
 star.  Its centrifugal support slows its collapse and the high
 angular momentum material does not immediately become part of the
 proto-neutron star.  As the proto-neutron star contracts further,
 this high angular momentum will not contract with it but will hang up
 in a disk around the proto-neutron star.  The high angular momentum
 material may well lose much of its angular momentum before accreting
 onto the neutron star through transport out the the disk or in a
 wind.  Recall, that it requires nearly $10^{51}$\,ergs of energy to
 eject $10^{-4}$\,M\sun on the surface of a 20\,km neutron star.  If
 the wind or disk outflows extract  rotation energy, much of the
 rotational energy may be lost in ejecting only a small amount of
 material off the disk surface.  It is much easier to transport
 angular momentum outwards than compress high angular momentum
 material onto the proto-neutron star.  Hence, the spin periods listed
 in Table 1 should be seen as {\it upper limits}.

Pulsars with the slow rotation periods predicted for our slowly 
rotating progenitor (SN15C) will not be bright enough to drastically 
change the supernova explosion energy, no matter what the magnetic 
field strength.  However, our fast rotating progenitors could easily 
produce 1-2\,ms pulsars that, with sufficiently high magnetic fields, 
could inject energies comparable to the supernova explosion 
itself.  A pulsar born with a 1\,ms period and a $10^{12}$\,G magnetic 
field would inject a $2\times10^{50}$erg jet during its first year.  
This is roughly 10-20\% of the total supernova energy and may be 
detectable in the supernova lightcurve.  A similar pulsar with 
with a $10^{13}$\,G magnetic would inject nearly $5\times10^{51}$erg 
in a jet.  This would dominate the supernova explosion energy.  It may 
be that the peculiar lightcurves and spectra of some supernova 
are exactly such objects:  fast spinning, high magnetic field 
pulsars.

We discuss pulsar emission and spin rates in more detail in 
\S 5.

\section{Other Rotational Observables:  Pulsars, Gravitational 
Waves and Nucleosynthesis}

In this paper, we have studied the collapse of a number of rotating
and non-rotating stars to probe the effects of rotation on supernovae.
We found that for our fastest rotating progenitor stars (the fastest
pre-collaapse produced by modern stellar evolution models), rotation
can indeed produce asymmetries asymmetries in the explosion.  However,
the cores these stars did not fragment in our collapse simulations and
analysis of the rotation suggest that the rotation rate must be much
higher than predicted by Heger et al. (2000) to cause the
proto-neutron star to fragment.  Even if magneto-hydrodynamic
instabilities are as efficient as Akiyama et al. (2003) claim, our
simulations do not lead to magnetic instabilities which dominate the
explosion.  But we can not rule out that they won't play any role in
the explosion mechanism.  Indeed, if the magnetic field of the
resultant pulsar is greater than $\sim 10^{12}$\,G, the pulsar
emission will alter the supernova explosion.  However, if the star is
slowly rotating, as predicted by Heger et al. (2003),
rotationally-driven effects will not occur and we can rule out
rotation as an important effect in core-collapse supernovae.
 
The explosion asymmetries from our fastest rotating stars provide an
easy explanation for the outward mixing of nickel and anomolous
$^{44}$Ti yields observed in supernova 1987A (Nagataki 2000, Hungerford 
et al. 2003).  Whether such high rotation velocities occur in stars 
is a crucial uncertainty in understanding the fate of massive stars.  
Unfortunately, stellar models are not yet sophisticated enough to 
say whether nature produces fast (Heger et al. 2000) or slow 
(Heger et al. 2003) rotating cores, and it is unlikely that stellar 
evolution models with accurate magnetic field and angular momentum 
transport algorithms will exist in the near future.

However, there are a number of other indirect methods by which might 
try to determine the rotation of the stellar core before collapse.  One is
the emission from young pulsars.  As we discussed in \S 4.3, a rapidly
spinning, high magnetic field, pulsar can emit enough energy to impact
the supernova explosion.  There is a large database of observed
pulsars which, if we could determine their spin evolution since birth,
would give us a clue about the birth spin rates of pulsar.  The
critical uncertainty here is determining how the spins of pulsars
evolve.  Figure 14 shows pulsar luminosities as a function of time for
a range of pulsar initial conditions and properties.  One of the key
uncertainties in such a calculation is the role of r-modes to spin
down the pulsar.  Recent results show that the r-mode amplitude
($\alpha_{\rm r-mode}$) will not exceed $10^{-3}$ (e.g. Schenk et
al. 2002) but this constraint strongly depends upon the neutron star
equation of state and the particular r-mode instability (e.g. Lee \&
Yoshida 2003).  To get a flavor of the range of results, we use the
pulsar evolution code developed in Fryer, Holz, \& Hughes (2002)
following the formalism of Ho \& Lai (2000).  We have run models with
r-mode maximum amplitudes ($\alpha_{\rm r-mode}$) of $10^{-3}$ and
$10^{-2}$ which we assume to {\it not} be dependent upon the
temperature evolution of the neutron star.  This overestimates the
strength of the true r-mode instability.  Below a maximum amplitude of
$10^{-3}$, r-modes play very little role in the pulsar luminosity.  By
1000 years, the effects of r-modes on the early spin evolution is 
hard to detect.  Likewise, a star initially rotating
at 10\,ms will appear very similar to a a 1\,ms initial rotating
pulsar within 10,000 years after the supernova explosion.  It would be
difficult to distinguish 200 years after the supernova explosion a
1\,ms pulsar whose magnetic field decays from $10^{13}$G down to
$10^{12}$G after 200\,years from a 10\,ms whose magnetic field
remained constant at $10^{12}$G for those 200\,years.

These similarities are due to the fact that pulsars lose most of their
angular momentum early in their evolution.  It takes less than
100\,years for a 1\,ms pulsar to spin down to a 2\,ms pulsar
(Fig. 15).  A $10^{13}$\,G, 1\,ms pulsar will have a period longer than
1\,s after 1 million\,years.  

Very few pulsars have age determinations that are independent of the
spin-down rate, so it is very hard to determine the true birth spin of
pulsars.  Estimates of the birth spin rates range from 2\,ms
(Middleditch et al. 2000) to well above 100\,ms (Romani \& Ng 2003).
However, all these estimates suffer from many uncertainties and it is
difficult to constrain stellar rotation rates from observations of
pulsar spins.  In general it is believed that some pulsars were
probably born with spin periods faster than 10\,ms, but some are also
born with much slower spin periods.  If some pulsars are born with
spin periods faster than 10\,ms, some stellar cores must be rotating
faster than our slowly rotating progenitor (SN15C).  If all pulsars
are born with spin rates above 20\,ms, our fast rotating models will
have to lose considerable angular momentum as they contract.  But 
there are models to spin up neutron stars and mechanisms to remove 
angular momentum, so it is unclear if pulsar observations can rule 
out any progenitor.

A more promising constraint on the initial rotating periods of massive 
stars is the gravitational wave signal.  In Fig. 16, we plot the 
gravitational wave signal (amplitudes) for model SN15B-nr and SN15B.  
The signal from the rotating progenitor can be over a factor of 5
higher than the non-rotating case.  Even so, the signal is not strong 
enough to be observable by advanced LIGO (see Gustafson et al. 1999 or 
Fryer, Holz, \& Hughes 2002).  But a galactic supernova would easily 
have a strong enough signal to distinguish our fast rotating models 
(SN15A,SN15B) with our slow rotating progenitor (SN15C) or non-rotating 
progenitors.  Indeed, it is likely that a galactic supernova will provide 
enough of a signal to distinguish rotating models from other asymmetries 
in the core-collapse (see Fryer, Holz, Hughes 2003 for more details).

Nuclear yields from supernova explosions could also place constraints
on the rotation.  The temperatures in the rotating core are roughly
symmetric (Fig. 17), slightly peaking in the polar region.  However,
high density contours extend much further out in the equator
(Fig. 6), so the neutrinosphere is deeper in along the poles than along
the equator.  These two effects both cause neutrinos arising from the
polar region to have higher mean energies than those in the equator.  
The neutrino-driven wind will be far from symmetric, and the r-process 
yields may differ considerably from that predicted by symmetric wind 
calculations (e.g. Terasawa et al. 2002, Thompson 2003).

\acknowledgements

We are grateful A. Heger and T. Mezzacappa for useful discussions and
support.  This work was funded under the auspices of the U.S. Dept. of
Energy, and supported by its contract W-7405-ENG-36 to Los Alamos
National Laboratory and by a DOE SciDAC grant number
DE-FC02-01ER41176.  The simulations were conducted both on the ASCI Q
machine and the Space Simulator, both at Los Alamos National
Laboratory.

\clearpage

\begin{figure}
\plotone{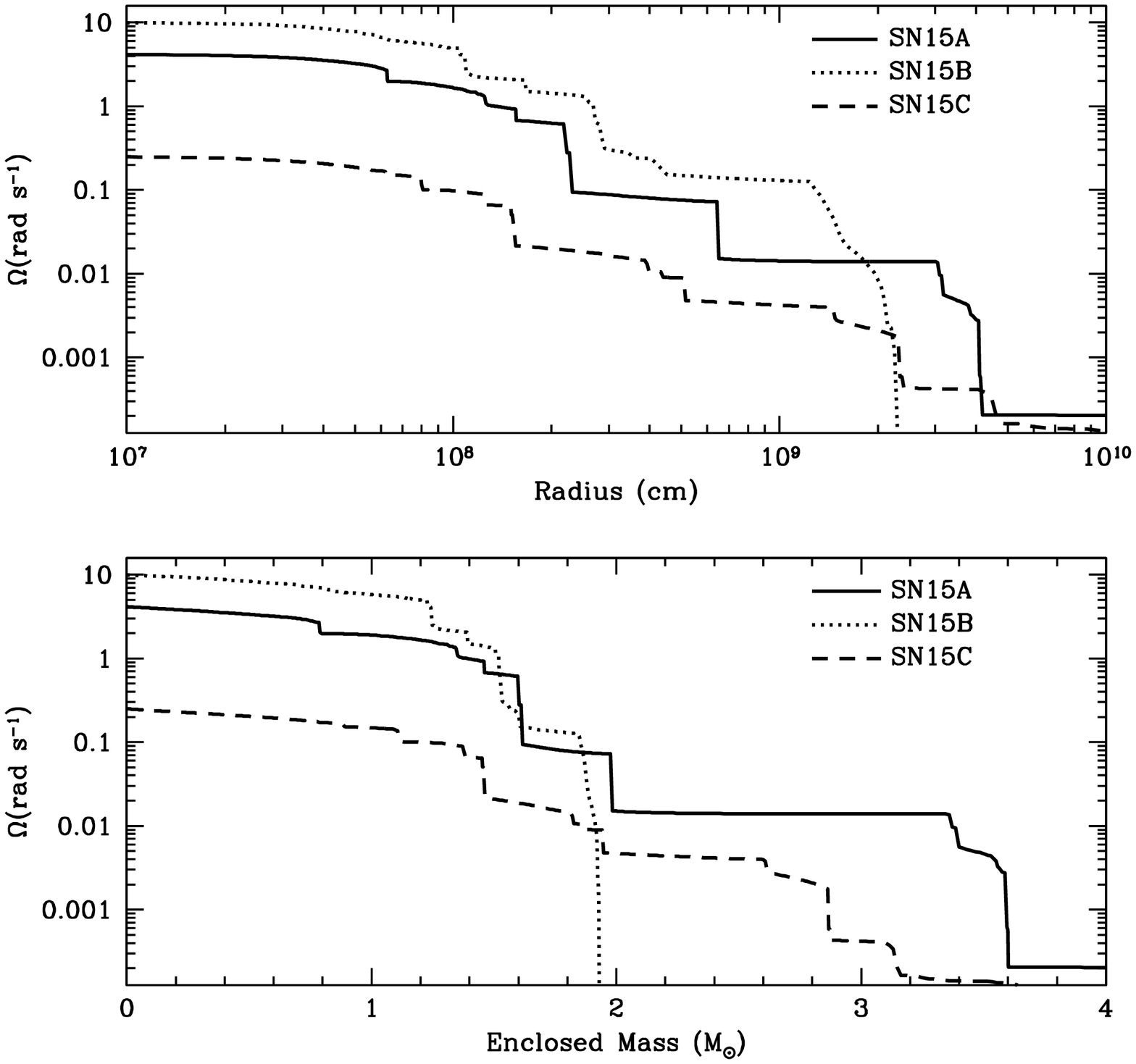}
\caption{Angular velocity versus radius (top) and mass (bottom) for 
our 3 basic progenitors:  SN15A, SN15B, (models E15A and E15B from 
Heger, Langer, \& Woosley 2000) and SN15C (15\,M\sun model from 
Heger, Woosley, \& Spruit 2003).  The angular velocity remains 
relatively constant in burning shells due to convection which 
efficiently transports angular momentum.  However, at the boundaries of 
these layers, the spins can decouple, causing jumps in the angular 
velocity.  These jumps persist, although with much smaller magnitudes, 
in the progenitor (SN15C) which includes magnetic fields (which can 
transport angular momentum across these boundaries). Note that 
FH erroneously plotted model SN15A instead of the model they used 
for their simulations (SN15B) in their Fig. 3.\label{fig1}}
\end{figure}

\begin{figure}
\caption{An angular slice (0.5$^{\circ}$) in the x-z plane of model 
SN15A-hr just before bounce.  By angular slice of 0.5$^{\circ}$, 
we mean:  $|y|/\sqrt{x^2+y^2+z^2} < {\rm sin}0.5^{\circ}$.  The 
colors denote radial velocity and the vectors denote velocity 
direction and magnitude (vector length).  The material in the 
equator (x-axis) is slowed by centrifugal forces and hence 
has a slower infall velocity than the material in the poles.\label{fig2}}
\end{figure}

\clearpage

\begin{figure}
\caption{Entropy isosurfaces (entropies of $7.5,8.0,9.0 k_{\rm B} 
{\rm \, per \, nucleon}$) for model SN15A-hr 45\,ms after bounce.  Because 
the velocities are much higher along the poles (Fig. 2), the shock is 
stronger, producing higher entropy material in this region after bounce.  
Hence, the highest entropy material is limited to the polar caps.\label{fig3}}
\end{figure}

\begin{figure}
\epsscale{1}
\plotone{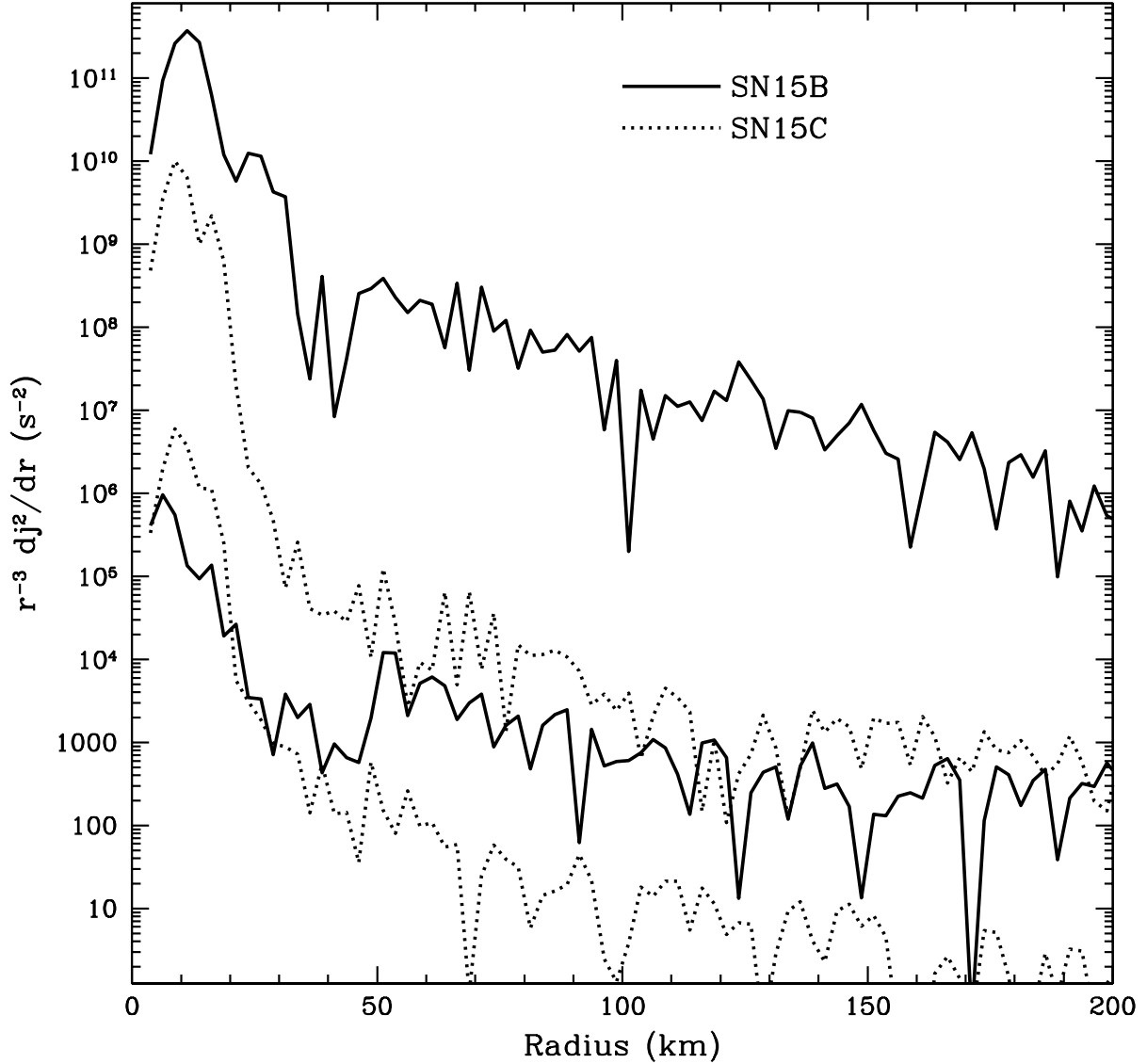}
\caption{The effect of angular momentum on convective instabilities
(second term in equation 3) for our fastest rotating model (SN15B -
solid lines) and our slowest rotating model (SN15C - dotted lines).
The upper curve is this term along the equator, the lower curve is 
this term along the rotation axis.  Not surprisingly, this value is 
low along the rotation axis.  Note also, that in the convection region, 
(beyond 50\,km), the value is 4 orders of magnitude lower in the slowly 
rotating model (SN15C) than in the fast rotating model (SN15B).  It 
is thus not surprising that rotation plays a much stronger role in the 
convection for model SN15B than in SN15C.  Indeed, in figure 5, we see 
that the convection in the slow rotating model compares better with 
the non-rotating cases than our fast rotating models (SN15A and SN15B).} 
\end{figure}

\clearpage

\begin{figure}
\caption{Upward moving material (isosurfaces of material moving with 
outward radial velocities of 3000\,km\,$s^{-1}$) for many of our 
models 45\,ms after bounce.  Note that the fast rotating models 
(SN15A, SN15A-hr, and SN15B) all convect most strongly along the poles, 
whereas the convection in the non-rotating or slowly rotating models 
(SN15A-nr, SN15B-nr, and SN15C) have no preferred direction. \label{fig5_a}}
\end{figure}

\begin{figure}
\epsscale{1}
\plotone{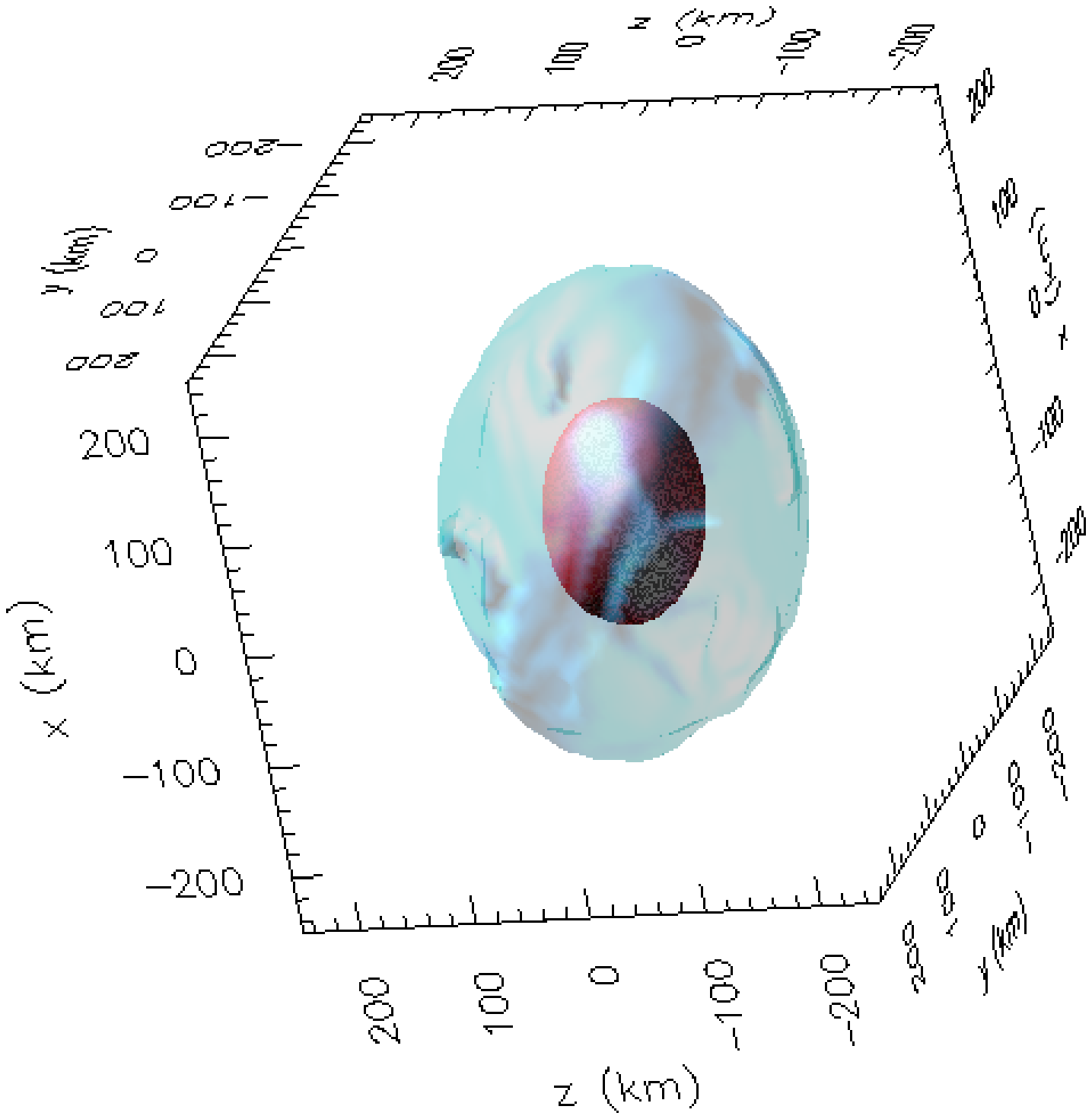}
\caption{Density isosurfaces ($10^{11}$ (blue), $10^{12}$ (red) ${\rm g \,
cm^{-3}}$) of model SN15B-hr 45\,ms after bounce.  Note that the density 
structure is asymmetric even at the compact structure of the high 
density ($10^{12} {\rm g \,cm^{-3}}$) isosurface.  Although the structure 
is asymmetric, the density remains centrally peaked (not toroidal), making 
it less easy to develop bar instabilities or fragment.
\label{fig6}}
\end{figure}

\clearpage

\begin{figure}
\caption{Upward moving material (isosurfaces of material moving with 
outward radial velocities of 1000\,km\,$s^{-1}$) for model SN15A 
20, 55, and 80\,ms after bounce.  The bulk of the upflows are constrained 
along the rotation axis, in agreement with the 2-dimensional results 
of FH. Ultimately, this convection will drive a stronger explosion 
along the poles.  \label{fig5_b}}
\end{figure}

\begin{figure}
\caption{An 0.5$^{\circ}$ angular slice (see Fig. 2 for details) 40\,ms 
after bounce of SN15A-hr.  The colors denote the specific angular momentum 
of the material and the vectors show velocity direction and magnitude 
(vector length).  Note that the bulk of the angular momentum lies along 
the equator (the angular momentum in the a $15^{\circ}$ cone along the 
poles is 2 orders of magnitude less than that in the equator).  The 
specific angular momentum in the equator is over $10^{16} {\rm 
cm^2 s^{-1}}$, corresponding to a rotation velocity of nearly $5000 
{\rm km s^{-1}}$ and a rotational period of 250\,ms.\label{fig7}}
\end{figure}

\clearpage

\begin{figure}
\plotone{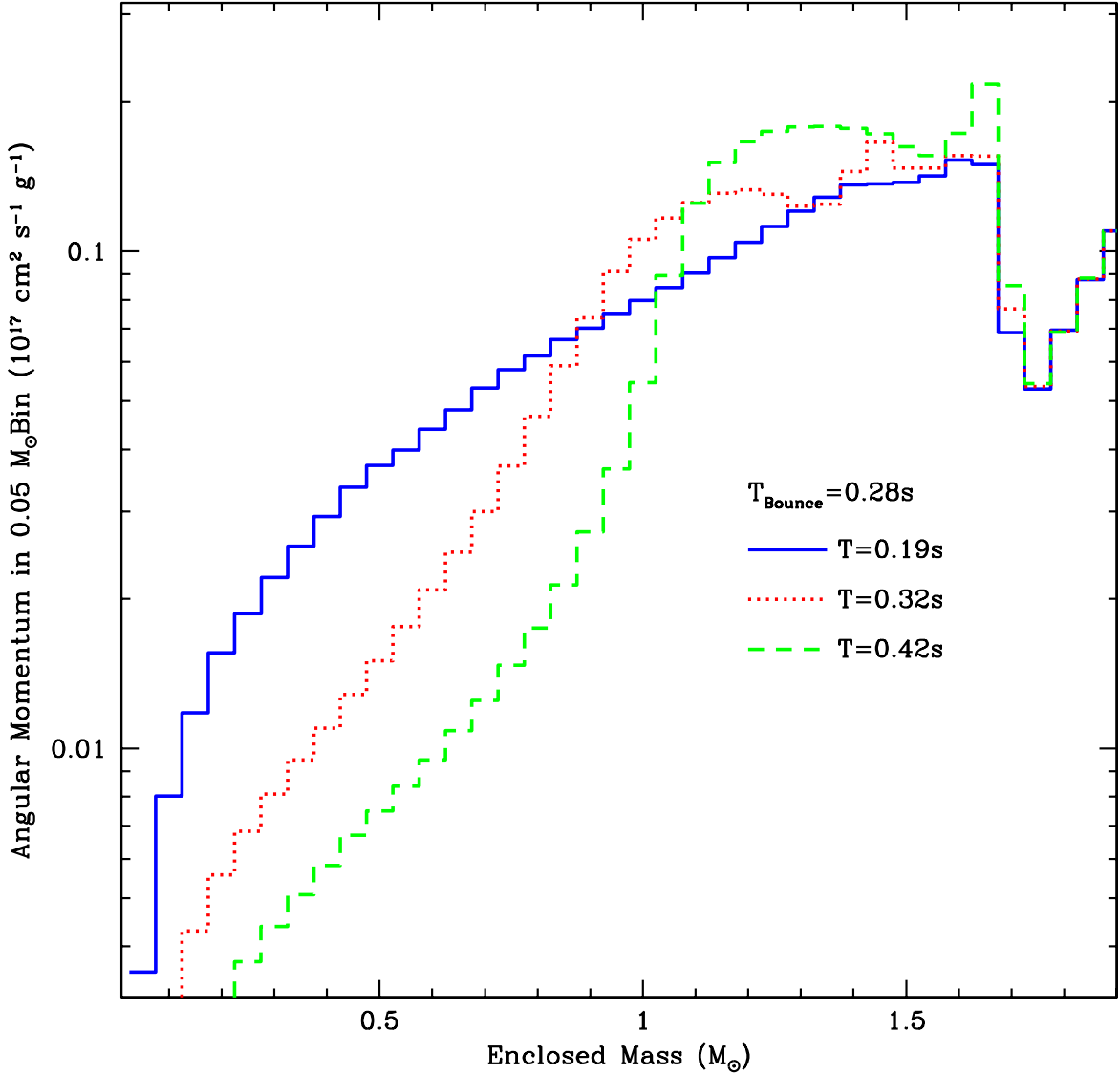}
\caption{Angular momentum versus mass zone as a function 
of time.  The solid line shows the angular momentum profile 90\,ms before 
bounce, the dotted line is 40\,ms after bounce, and the dashed line is 
140\,ms after bounce.  Note that in the proto-neutron star interior, the 
star quickly loses 80\% of its total angular momentum.  This angular momentum 
is transported to the surface of the proto-neutron star (note the rise 
in angular momentum beyond 1\,M\sun 140\,ms after bounce).  \label{fig8}}
\end{figure}

\begin{figure}
\plotone{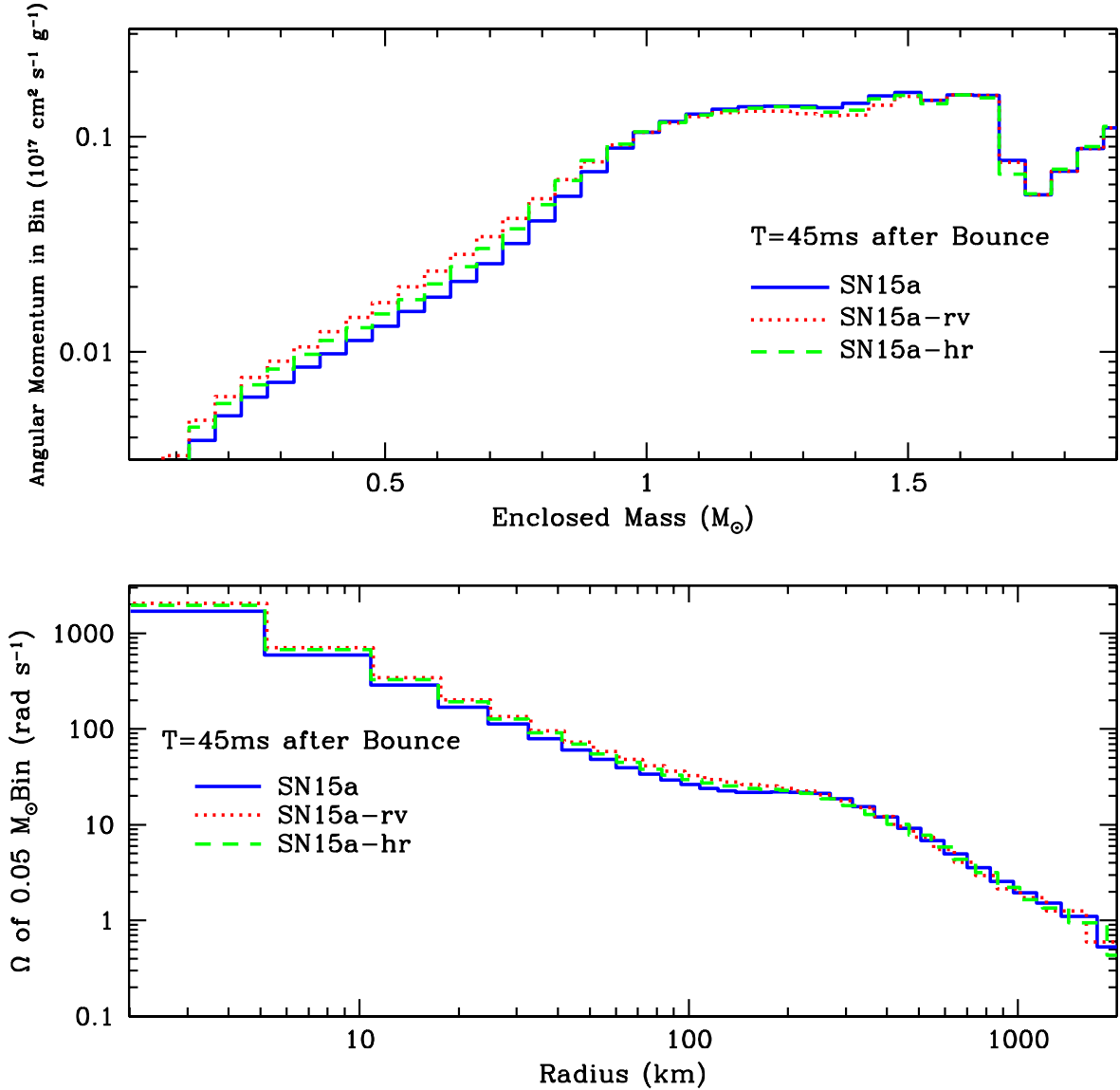}
\caption{Angular momentum versus mass (top) and radius (bottom) for 3 
separate models (SN15a:  solid line, SN15a-rv:  dotted line, SN15-hr:  
dashed line).  From Fig. 8, we note that by 45\,ms after bounce, considerable 
transport has already occured.  However, the differences between the 
high resolution (SN15A-hr), reduced viscosity (SN15A-rv), and standard 
(SN15A) models are small.  If the angular momentum transport were numerical, 
one would expect the high resolution and reduced viscosity runs would have 
different results.  Although slightly less transport has occured in the 
reduced viscosity run, there is not a factor of 10 difference which one 
would expect by decreasing the viscosity by a factor of 10.  We can 
be reasonably assured that the transport is not a numerical artifact.  
\label{fig9}}
\end{figure}

\clearpage

\begin{figure}
\plotone{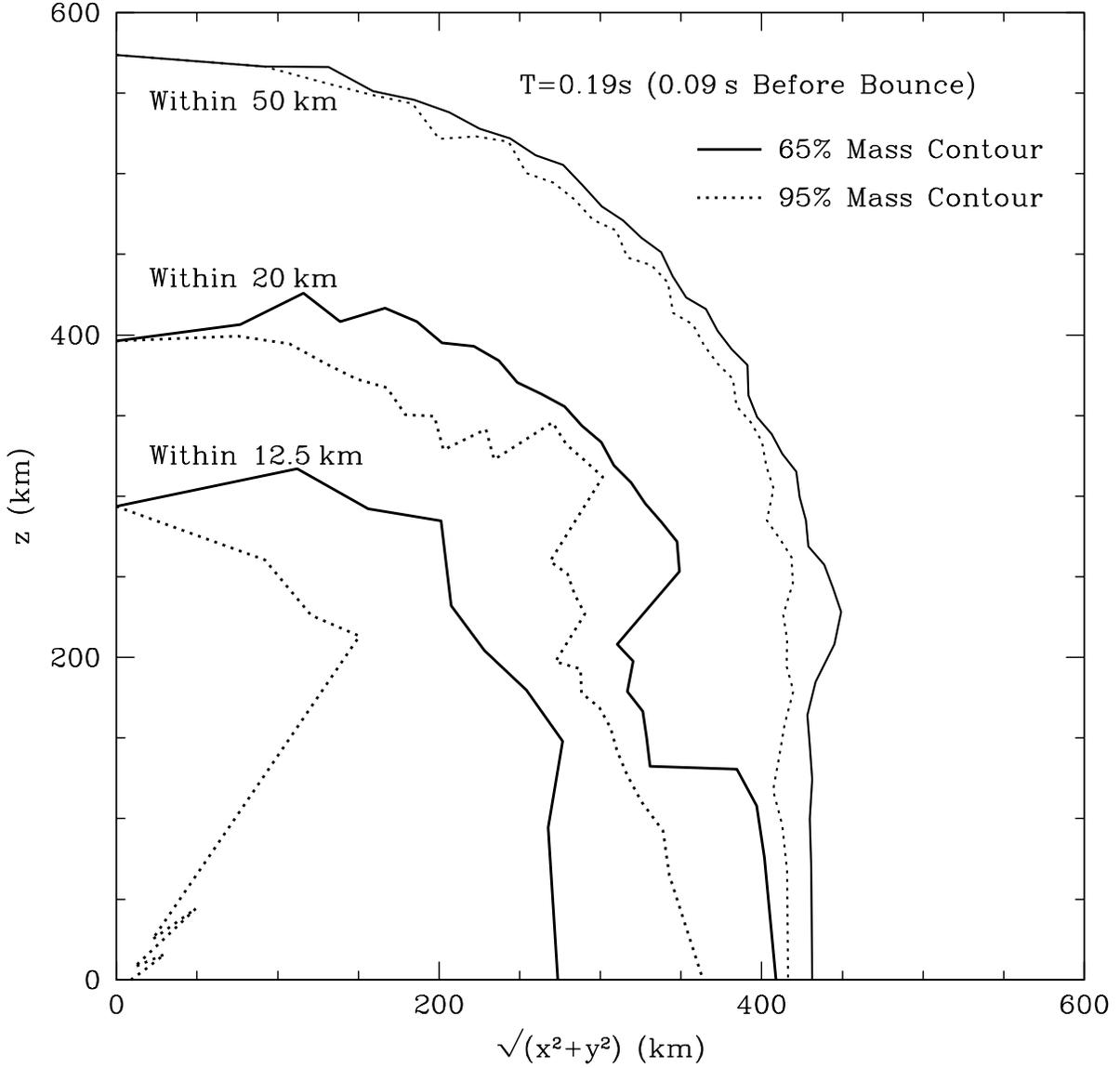}
\caption{Distribution of material in model SN15A 90\,ms before bounce
which ultimately contracts to form the proto-neutron star.  A large
fraction of the material below the isocontour boundaries collapse down
to 12.5, 20, 50\,km 140\,ms after bounce.  The fraction of material at
these boundaries which actually collapses down to 12.5, 20, or 50\,km
140\,ms after bounce is denoted by solid (65\%) and dotted (95\%)
lines.  This figure shows us what part of the original star collapses
within 12.5, 20, and 50\,km.  It is this material that ultimately forms
the neutron star.  Material in the equator does not contract so
quickly and does not initially become part of the neutron star and the
neutron star mass is originally biased toward the low angular momentum
material of the polar region.\label{fig10}}
\end{figure}

\begin{figure}
\plotone{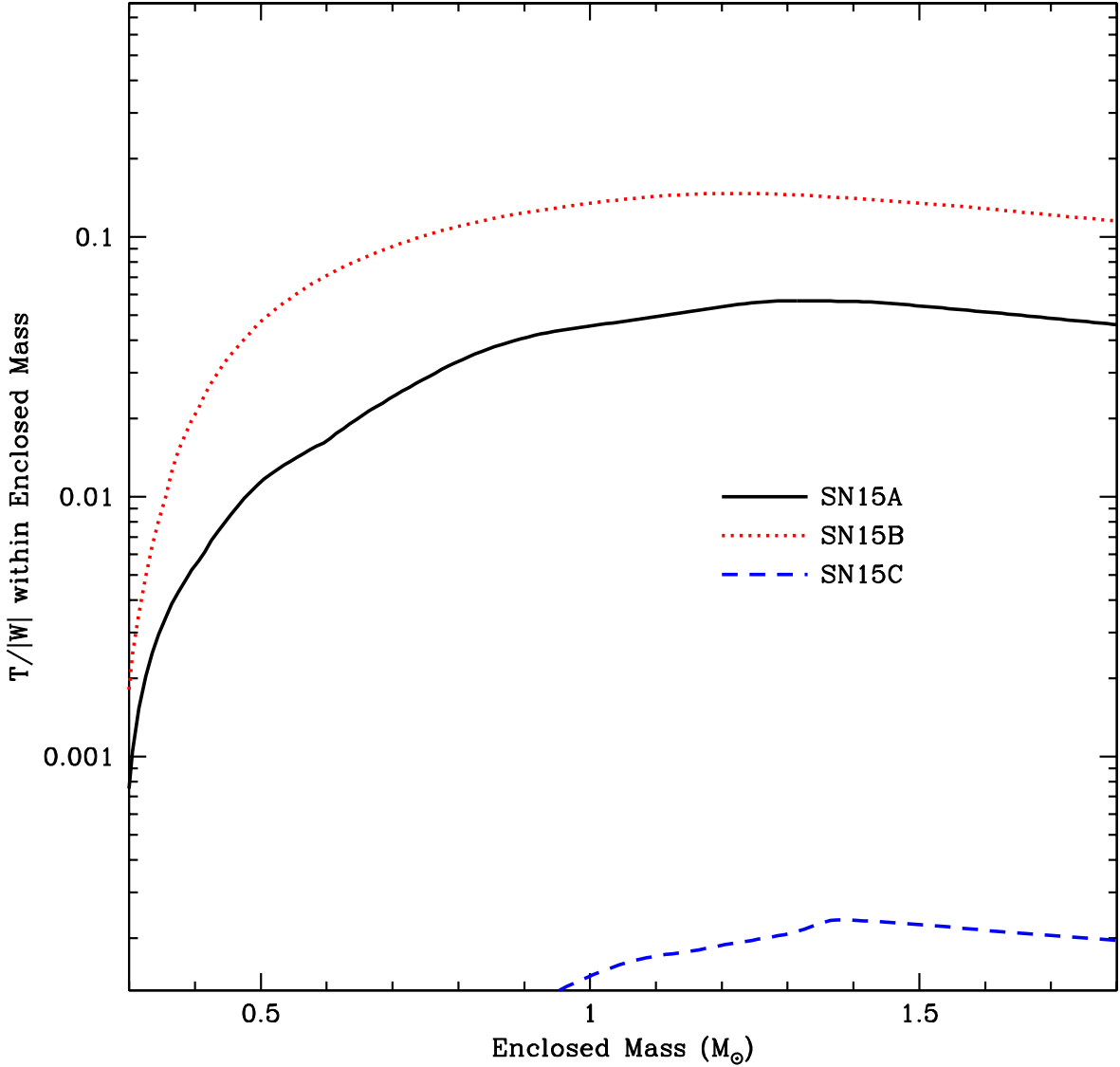}
\caption{The summation of the ratio of kinetic energy over potential
energy $T/|W|$ versus enclosed mass for our 3 rotating models: SN15A
(solid), SN15B (dotted), and SN15C (dashed).  For centrally peaked 
density profiles of our collapse models, this ratio must be above 
$\sim 0.25$ to produce dynamical instabilities.  For secular instabilities, 
this critical value is only $\sim 0.14$ and model SN15B slightly exceeds 
this value (it peaks at 0.147), but it is unlikely that secular 
instabilities will lead to fragmentation.  It is highly improbable that 
fragmentation will occur for most core-collapse supernovae.  Indeed, 
from this figure we see that it is unlikely that core-collapse 
stars will have any bar instabilities.
\label{fig11}}
\end{figure}

\clearpage

\begin{figure}
\plotone{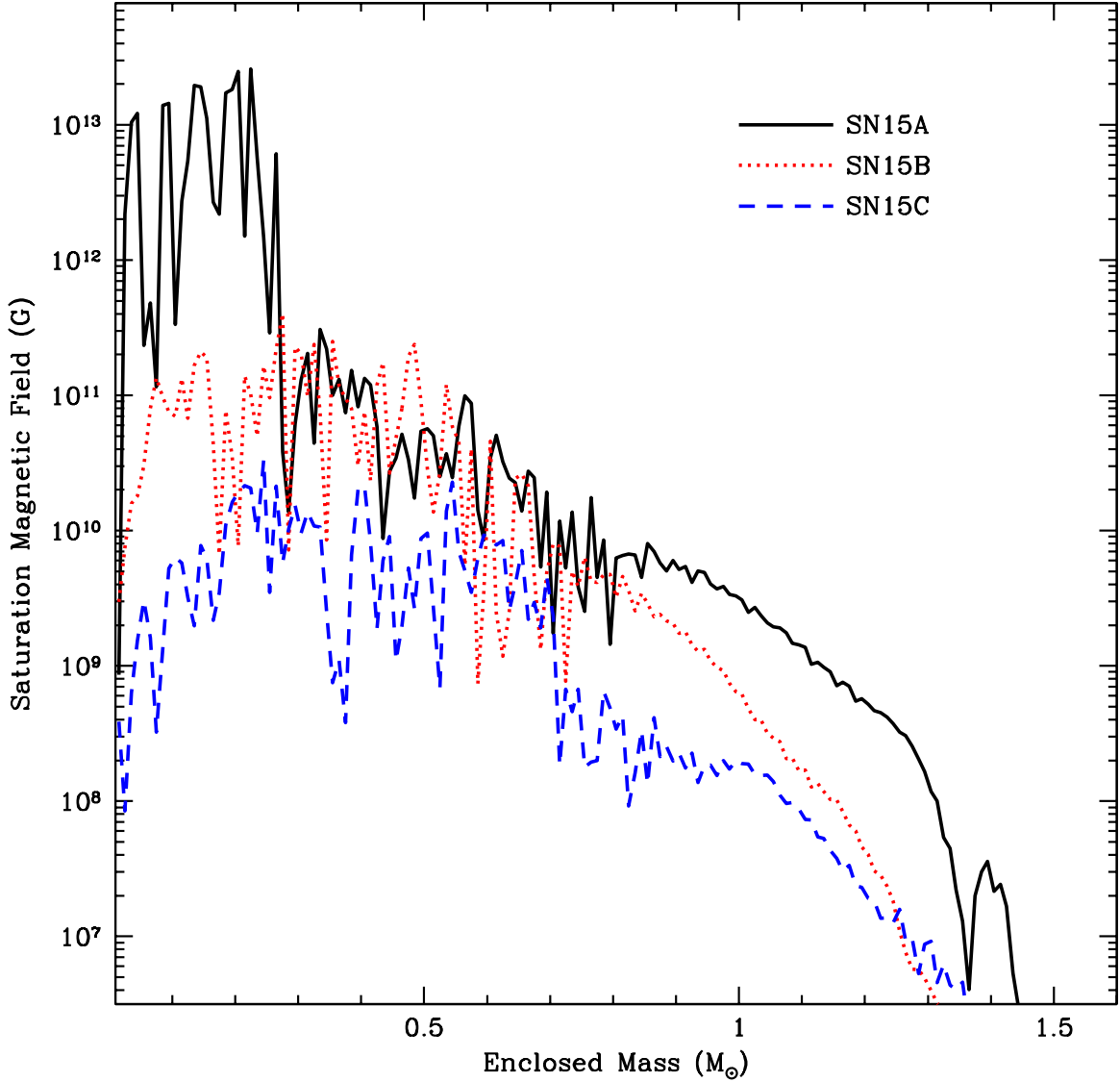}
\caption{Saturation magnetic field strength vs. enclosed mass for our
3 rotating progenitors (SN15A, SN15B, SN15C).  Note that the magnetic
fields are much smaller than that predicted by Akiyama et al. (2003)
and never exceed $10^{14}$G.  Such weak magnetic fields will not
dominate the explosion mechanism, but may alter the convection 
enough to drive asymmetries.  However, the strength of the magnetic 
field will depend strongly on how compressed the proto-neutron star 
can become.  At later times, or with a different equation of state 
for nuclear matter, magnetic fields may become more important.
\label{fig13}}
\end{figure}

\begin{figure}
\plotone{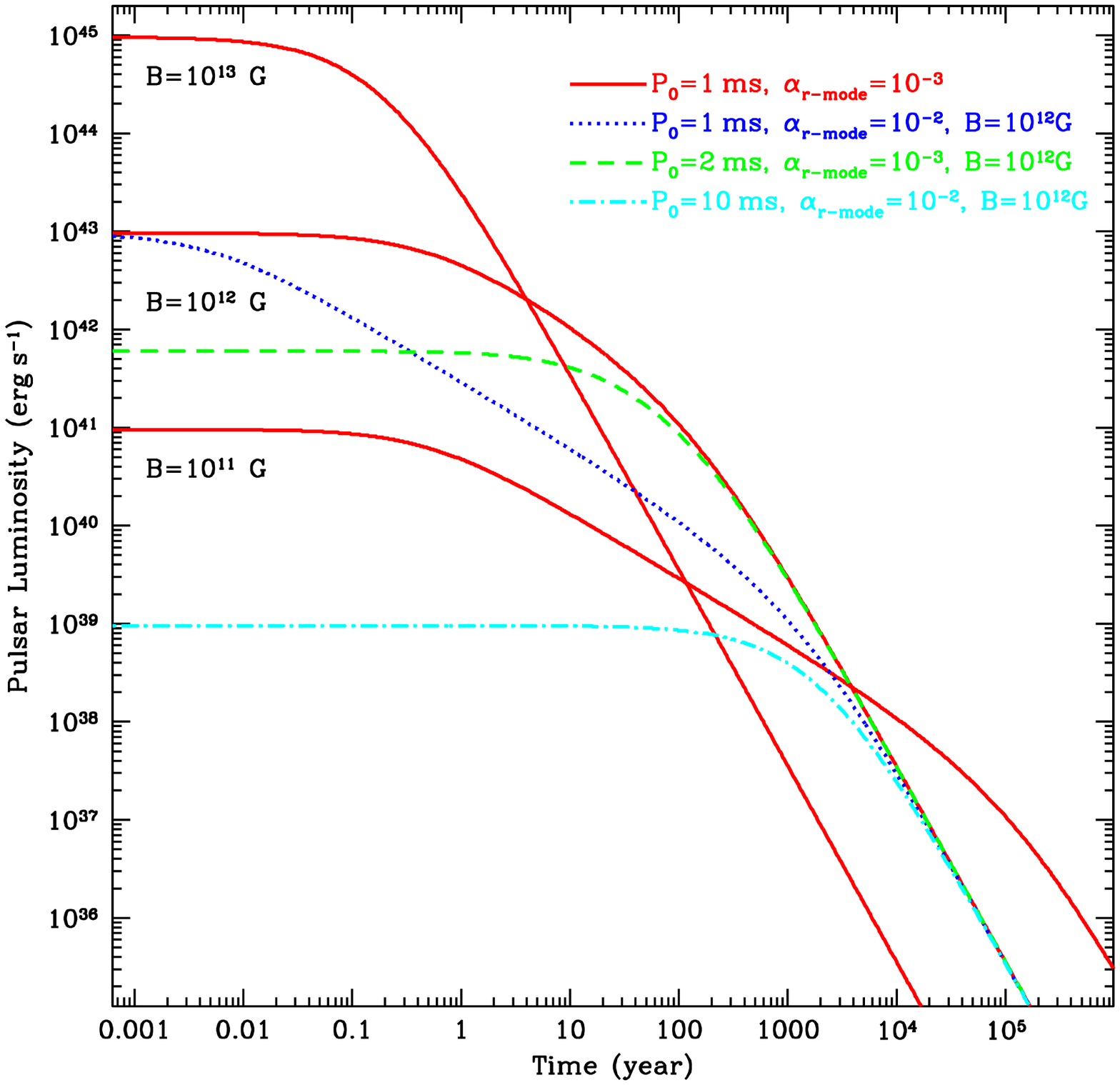}
\caption{Pulsar luminosity vs. time since formation.  The solid lines 
denote pulsars with initial spin periods of 1\,ms, maximum r-mode 
amplitudes:  $\alpha_{\rm r-mode}=10^{-3}$, and a range of magnetic 
field strengths ($10^{11},10^{12},10^{13}$\,G).  The dotted,dashed lines show 
the evolution of a $10^{12}$\,G field neutron star with, respectively, 
a higher maximum r-mode amplitude: $\alpha_{\rm r-mode}=10^{-2}$ and 
a 2\,ms initial period.  Note that after 1000\,years, it is very hard 
to distinguish different the initial structure of the star by its 
luminosity.
\label{fig14}}
\end{figure}

\clearpage

\begin{figure}
\plotone{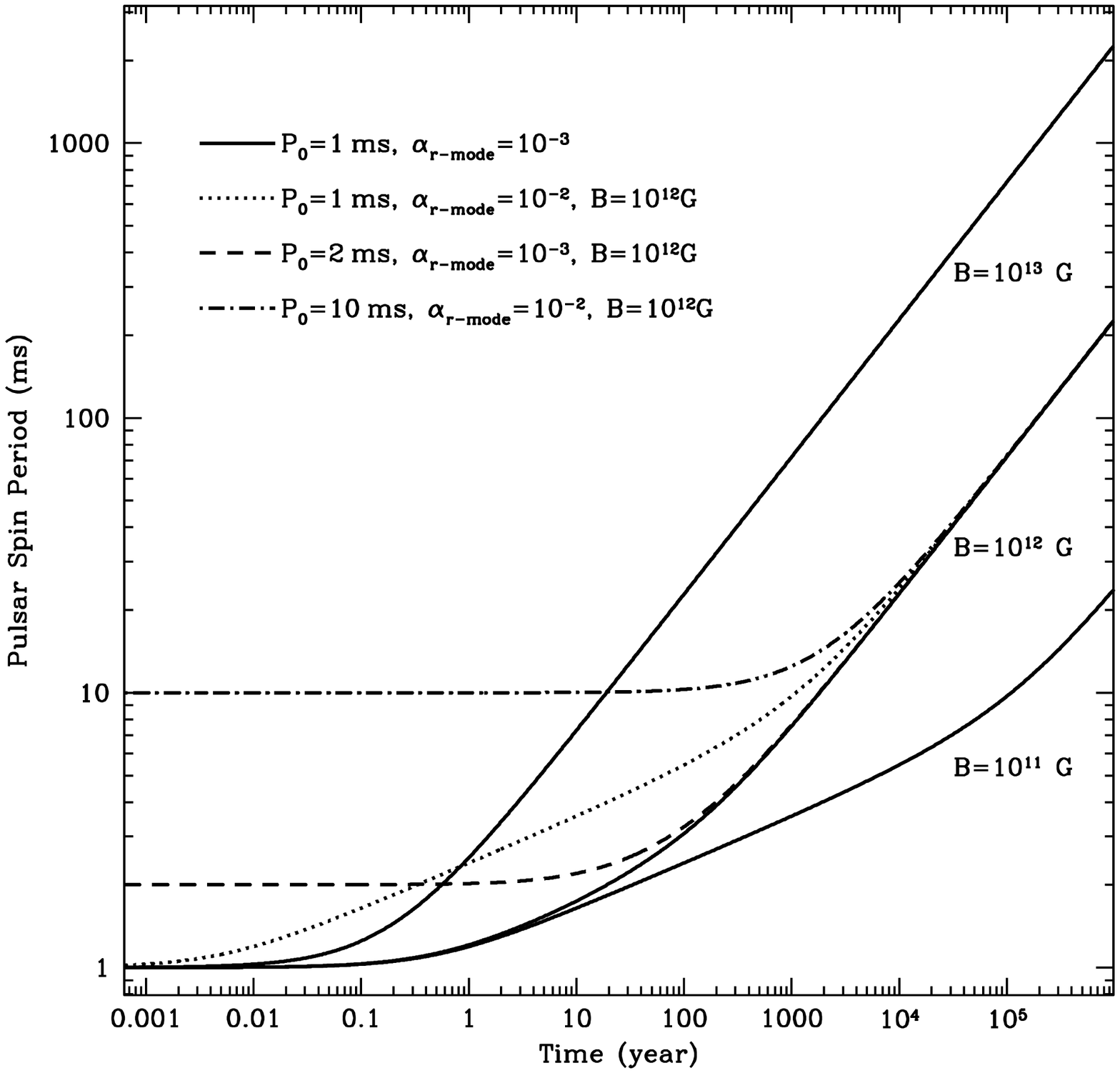}
\caption{Pulsar spin vs. time since formation.  The solid lines 
denote pulsars with initial spin periods of 1\,ms, maximum r-mode 
amplitudes:  $\alpha_{\rm r-mode}=10^{-3}$, and a range of magnetic 
field strengths ($10^{11},10^{12},10^{13}$\,G).  The dotted,dashed lines show 
the evolution of a $10^{12}$\,G field neutron star with, respectively, 
a higher maximum r-mode amplitude: $\alpha_{\rm r-mode}=10^{-2}$ and 
a 2\,ms initial period.  Note that after 1000\,years, it is very hard 
to distinguish different the initial structure of the star by its 
spin. \label{fig15}}
\end{figure}

\begin{figure}
\plotone{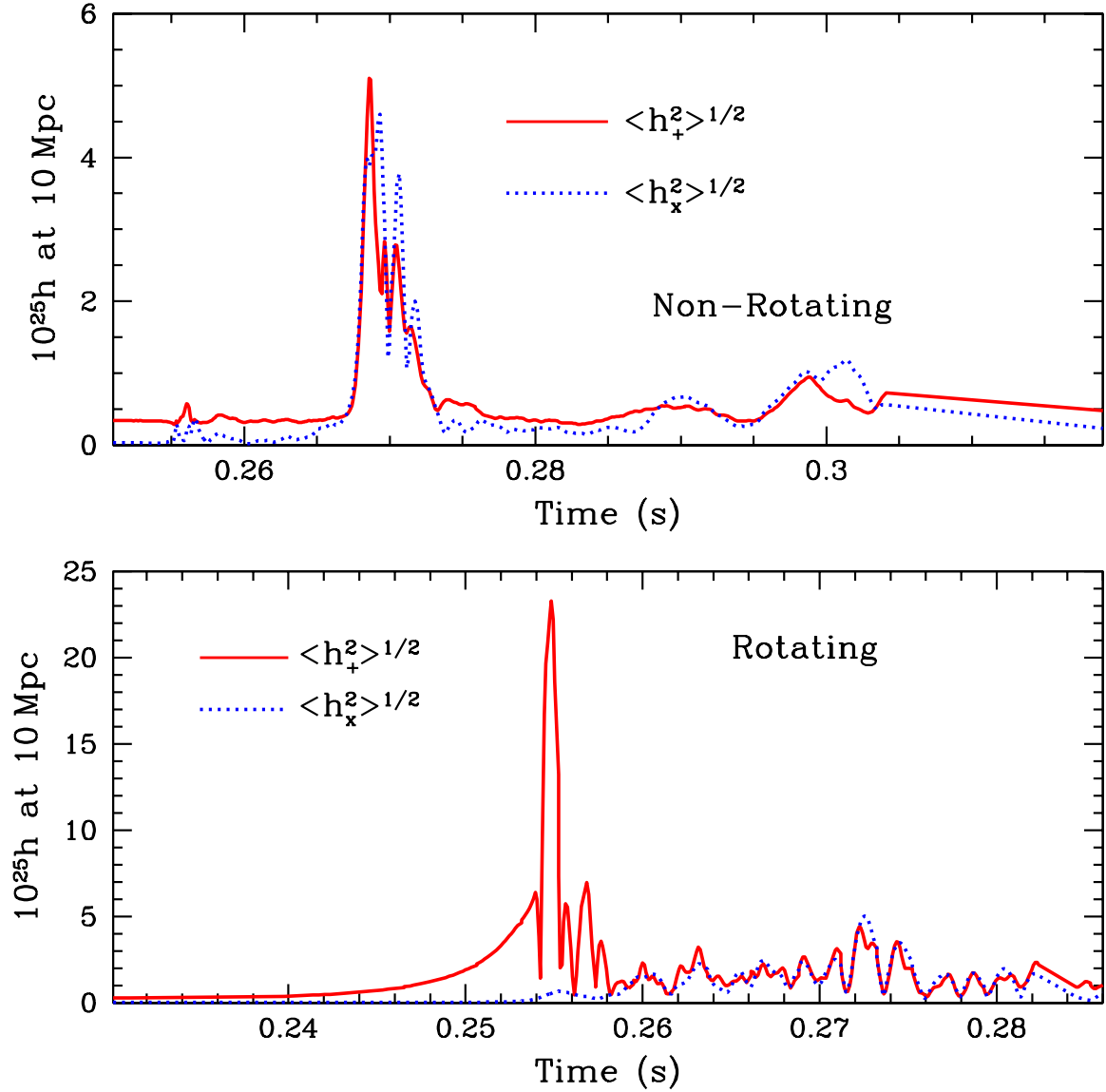}
\caption{The angle averaged wave amplitudes ($<h^2_{+}>^{1/2}$: solid
line,$<h^2_{x}>^{1/2}$: dotted line) for models SN15B-nr (top) and
SN15B (bottom) versus time.  Note that the gravitational wave signal 
is a factor of 5 higher for the rotating model.  A galactic rotating 
supernova would be detectable by advanced LIGO.
\label{fig16}}
\end{figure}

\clearpage

\begin{figure}
\plotone{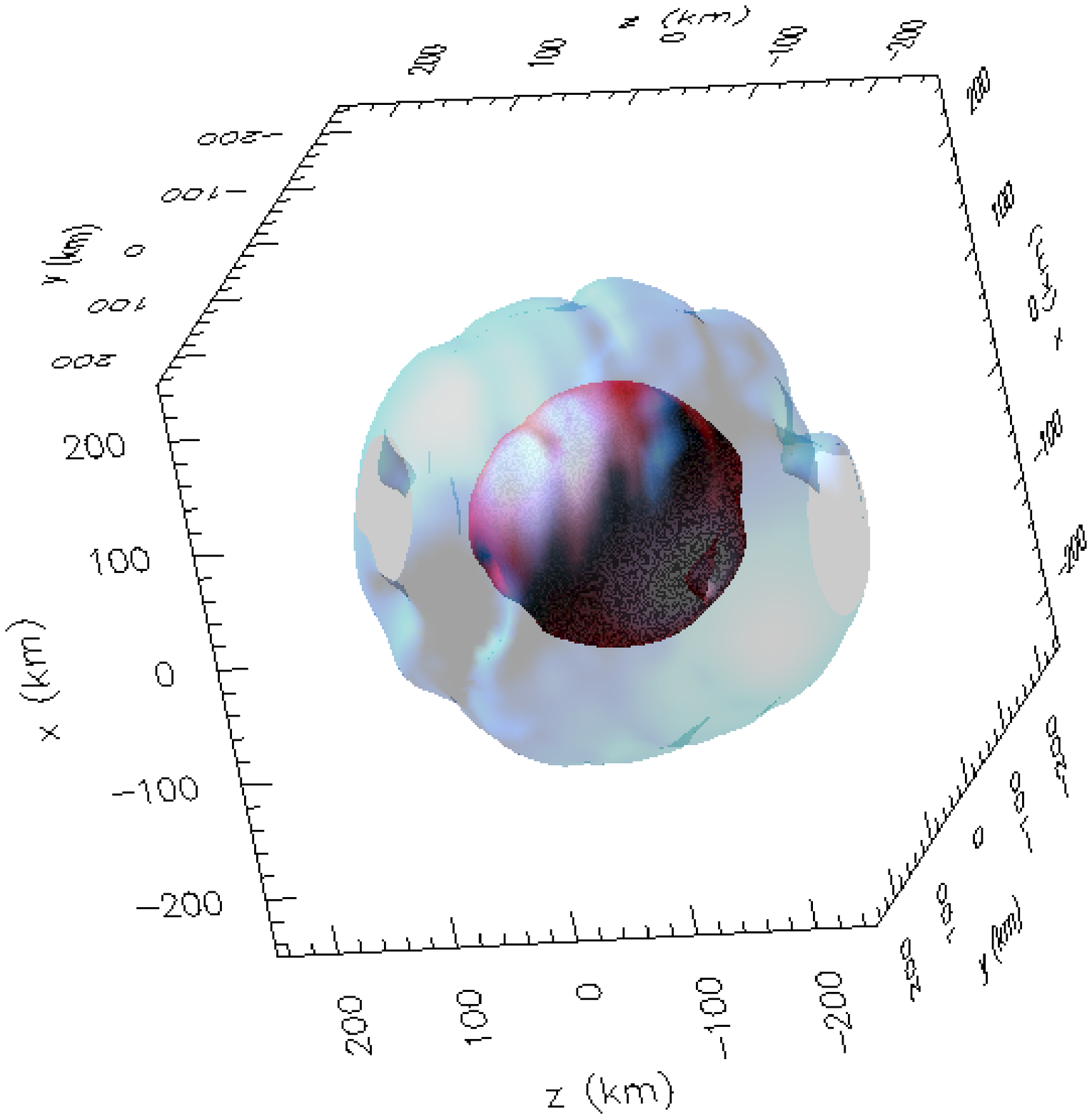}
\caption{Temperature isosurfaces ($2\times10^{10}$ (blue),
$3.5\times10^{10}$ (red) K) of model SN15B-hr 45\,ms after bounce.  Note
that the temperatures are only slightly asymmetric (with peaks in the 
polar regions).  The lower optical depth in the polar regions (see 
density profiles in Fig. 6) along with the higher temperatures in 
the poles lead to a hotter neutrino flux along the poles, and hence 
a higher entropy outflow.
\label{fig17}}
\end{figure}

\clearpage

\begin{deluxetable}{lccccccc}
\tabletypesize{\scriptsize}
\tablecaption{Core-Collapse Simulations \label{tbl-1}}
\tablewidth{0pt}
\tablehead{
\colhead{Model} & \colhead{Resolution} 
& \colhead{$P_{\rm core}^{\rm before collapse}$}\tablenotemark{a}
& \colhead{$\rho_{\rm bounce}$}\tablenotemark{b}
& \colhead{$E_{\rm rot}^{\rm 45\,ms}$}\tablenotemark{c}
& \colhead{$T/|W|^{\rm 45\,ms}$}\tablenotemark{d}
& \colhead{$B_{\rm max}^{\rm 45\,ms}$}\tablenotemark{e}
& \colhead{$P_{\rm NS}$}\tablenotemark{f} \\
\colhead{} & \colhead{(No. of Particles)} & \colhead{(s)} 
& \colhead{($10^{14}$g\,cm$^{-3}$)} & \colhead{($10^{51}$ ergs)} & \colhead{} 
& \colhead{(G)} & \colhead{(ms)}}
\startdata

SN15A & 1 million & 1.5 & 2.8 & 5(2) & 0.056 (0.035) & $10^{13}$ & 0.65 (0.91) \\
SN15A-hr & 5 million & 1.5 & 2.3 & 4 & 0.052 & $10^{13}$ & 0.66 \\
SN15A-rv & 1 million & 1.5 & 2.8 & 6 & 0.062 & $10^{13}$ & 0.64 \\
SN15A-nr & 1 million & 0 & 3.5 & $<0.001$ & $<10^{-4}$ & $<10^9$ & $>1000$ \\
SN15B & 0.5 million & 0.63 & 2.0 & 7 & 0.156 & $10^{11}$ & 0.35 \\
SN15B-nr & 1 million & 0 & 3.5 & $<0.001$ & $<10^{-4}$ & $<10^9$ & $>1000$ \\
SN15C & 1 million & 25 & 3.4 & 0.02 (0.02) & $2.3 \times 10^{-4}$ 
($2.0 \times 10^{-4}$) & $<10^{10}$ & 12 (17) \\

\enddata

\tablenotetext{a}{Rotation Period of the inner core from Heger et
al. (2002,2003) at collapse.}
\tablenotetext{b}{Density when the core ``bounces'' and drives a shock
out through the star.}
\tablenotetext{c}{Rotational energy of the inner 1.2\,M\sun, 45\,ms
after bounce.  We have also included, in parantheses, the rotational
energy for models SN15A 160\,ms after bounce and SN15C 135\,ms after
bounce.}
\tablenotetext{d}{The maximum of the ratio of kinetic to the absolute
value of the gravitational potential energy 45\,ms after bounce.  This
value should be above $\sim 0.14$ to have any chance of driving
significant bar instabilities.  As with the rotation energy, the
paranthetical values are for models SN15A 160\,ms after bounce and
SN15C 135\,ms after bounce.}
\tablenotetext{e}{Saturation magnetic field strengths using equation 
5 45\,ms after bounce.}
\tablenotetext{f}{Pulsar spin periods assuming that the angular
momentum in the inner 1.0\,M\sun is conserved as the star collapses
down to to a neutron star.  The paranthetical values are for models
SN15A 160\,ms after bounce and SN15C 135\,ms after bounce.  Note 
that the inner core continues to lose angular momentum.  Model SN15C 
is unlikely to produce pulsars with spin periods faster than 20\,ms. 
and it is unlikely that sub-millisecond pulsar will form from any 
model.}

\end{deluxetable}


\begin{thebibliography}{}

\bibitem[Akiyama et al. (2003)]{Aki03} Akiyama, S., Wheeler, J. C., 
Meier, D. L., \& Lichtenstadt, I. 2003, ApJ, 584, 954
\bibitem[Bazan \& Arnett (1998)]{Baz98} Bazan, G., \& Arnett, D. 1998, 
ApJ, 496, 316
\bibitem[]{Benz90} Benz, W. 1990, in The Numerical Modeling of Nonlinear 
Stellar Pulsations, ed. J. R. Buchler, 269
\bibitem[]{Bru01} Bruenn, S. W.; De Nisco, K. R.; Mezzacappa, A 2001, 
ApJ, 560, 326
\bibitem[]{Bur03} Buras, R.; Rampp, M.; Janka, H.-Th.; Kifonidis, K. 
2003, PRL, 90, 241101
\bibitem[Burrows et al.(1995)]{Bur95} Burrows, A., Hayes, J., \& Fryxell, 
B. A. 1995, ApJ, 450, 830
\bibitem[Colpi \& Wasserman 2002]{Col02} Colpi, M., \& Wasserman, I. 
2002, ApJ, 581, 1271
\bibitem[Davies et al.(2002)]{Dav02} Davies, M.B., King, A., Rosswog, 
S., \& Wynn, G. 2002, ApJ, 579, L63
\bibitem[]{Dim02} Dimmelmeier, H., Font, J. A., \& M\"uller, E. 2002, 
A\&A, 393, 523
\bibitem[1978]{End78} Endal, A. S., Sofia, s. 1978, ApJ, 220, 279
\bibitem[Fryer(1999)]{Fry99} Fryer, C. L. 1999, ApJ, 522, 413
\bibitem[FH]{Fry00} Fryer, C. L. \& Heger, A. 2000, ApJ, 541, 1033 - FH
\bibitem[Fryer, Woosley, \& Heger (2001)]{Fry01} Fryer, C. L., Woosley, 
S. E., \& Heger, A. 2001, ApJ, 550, 372
\bibitem[FK]{FK01} Fryer, C. L. \& Kalogera, V. 2001, ApJ, 554, 548
\bibitem[]{Fry02_a} Fryer, C. L., Holz, D. E., \& Hughes, S. A. 2002, 
ApJ, 565, 430
\bibitem[]{FW02} Fryer, C. L. \& Warren, M. S. 2002, ApJ, 574, L65
\bibitem[]{Fry02_b} Fryer, C. L., Holz, D. E., \& Hughes, S. A. 2003, 
in preparation
\bibitem[Gustafson et al. (1999)]{whitepaper} Gustafson, E., Shoemaker, D.,
Strain, K., Weiss, R. 1999 LSC White Paper on Detector Research and
Development, LIGO Document T990080-00-D.
\bibitem[]{Heg00} Heger, A., Langer, N., \& Woosley, S. E. 2000, ApJ, 
528, 368
\bibitem[]{Heg03} Heger, A., Woosley, S. E., \& Spruit, H. 2003, 
in preparation
\bibitem[Herant et al.(1994)]{Heretal94} Herant, M.,  Benz, W., Hix, W.R., 
Fryer, C.L. \& Colgate, S.A. 1994, ApJ, 435, 339
\bibitem[Ho \& Lai (2000)]{Ho00} Ho, W. C. G., \& Lai, D. 2000, ApJ, 
543, 386
\bibitem[H\"oflich et al. (1996)]{Hof96} H\"oflich, P., Wheeler, J. C., 
Hines, D. C., \& Trammell, S. R. 1996, SpJ, 459, 307
\bibitem[Hungerford et al. (2003)]{Hun03} Hungerford, A. L., Fryer, 
C. L., \& Warren, M. S. 2003, accepted by ApJ
\bibitem[Iwamoto et al. 1998]{Iwa98} Iwamoto, K. et al. 1998, 
Nature, 395, 672
\bibitem[Janka \& M\"uller(1996)]{Jan96} Janka, H.-Th., \& M\"uller, E. 
1996, A\&A, 306, 167 
\bibitem[Kotake et al. 2003]{Kat03} Kotake, K., Yamada, S., \& 
Sato, K. 2003, accepted by ApJ
\bibitem[Lattimer \& Swesty(1991)]{Lat91}
Lattimer, J.M., \& Swesty, F.D. 1991, Nuc. Phys. A, 535,331
\bibitem[Leblanc \& Wilson (1970)]{LeB70} LeBlanc, J. M. \& Wilson, 
J. R. 1970, ApJ, 161, 541
\bibitem[Lee \& Yoshida (2003)]{Umi03} Lee, U., \& Yoshida, S. 2003,
ApJ, 586, L403
\bibitem[]{Lie01} Liebend\"orfer, M., Mezzacappa, A., 
Thielemann, F.-K., Messer, O. E., Hix, W. R., Bruenn, S. W. 2001, 
Phys. Rev. D, 63, 103004
\bibitem[]{Mez93} Mezzacappa, A., \& Bruenn, S. W. 1993, ApJ, 410, 740
\bibitem[Middleditch et al. (2000)]{Mid00} Middleditch, J. et al. 
2000, New Astronomy, 5 243
\bibitem[M\"uller \& Hillebrandt (1981)]{Mul81} M\"uller, E. \& 
Hillebrandt, W. 1981, A\&A, 103, 358
\bibitem[Nagataki (2000)]{Nag00} Nagataki, S. 2000, ApJS, 127, 141
\bibitem[Centrella et al. (2001)]{Cen01} Centrella, J. M., New, 
K. C. B., Lowe, L. L., \& Brown, J. D. 2001, ApJ, 550, L193 
\bibitem[New (2003)]{Nag03} New, K. C. B. 2003, Living Riviews in 
Relativity, 6, 2
\bibitem[Rampp \& Janka (2002)]{Ram02} Rampp, M., \& Janka, H.-T. 
2002, A\&A, 396, 361
\bibitem[Romani \& Ng (2003)]{Rom03} Romani, R. W., \& Ng, C.-Y. 
2003, ApJ, 585, L41
\bibitem[Schenk et al. 2002]{Sch02} Schenk, A. K., Arras, P., Flanagan, 
\'E. \'E., Teukolsky, S. A., and Wasserman, I. 2002, PRD, 65, 4001
\bibitem[Shimizu et al. 1994]{Shi94} Shimizu, T., Yamada, S., \& 
Sato, K. 1994, ApJ, 432, L119
\bibitem[]{Ter02} Terasawa, M., Sumiyoshi, K., Yamada, S., Suzuki, H., 
\& Kajino, T. 2002, ApJ, 578, L137
\bibitem[]{Tho03} thompson, T. A. 2003, ApJ, 585, L33
\bibitem[Wang (2002)]{Wan02} Wang et al. 2002, ApJ, 579, 671
\bibitem[Warren \& Salmon(1993)]{War93} Warren, M.S., \& Salmon, J.K. 
1993, In Supercomputing '93, 12-21, 1993 IEEE Comp. Soc.
\bibitem[Warren \& Salmon(1995)]{War95} Warren, M.S., \& Salmon, J.K. 
1995, Computer Physics Communications, 87, 266
\bibitem[Warren, Rockefeller, \& Fryer(2002)]{War02} Warren, M.S., 
Rockefeller, G.,  \& Fryer, C.L. 2003, in preparation
\bibitem[Woosley (1993)]{Woo93} Woosley, S. E. 1993, ApJ, 405, 273

\end{thebibliography}
\end{document}